\documentclass[twocolumn,showpacs,preprintnumbers,amsmath,amssymb]{revtex4}
\usepackage{graphicx}% Include figure files
\usepackage{dcolumn}% Align table columns on decimal point
\usepackage{bm}% bold math

\begin{document}

\preprint{APS/XXX-YYY}

\title{Resonant Kelvin-Helmholtz modes in sheared relativistic flows}

\author{Manuel Perucho$^{1,2}$}\email{perucho@mpifr-bonn.mpg.de}
\author{Michal Hanasz$^3$}
\author{Jos\'e-Mar\'{\i}a Mart\'{\i}$^1$}
\author{Juan-Antonio Miralles$^4$}
\affiliation{$^1$Departament d'Astronomia i
Astrof\'{\i}sica,
Universitat de Val\`encia \\
$^2$Max-Planck-Institut f\"ur Radioastronomie, Bonn\\
$^3$Toru\'n Centre for Astronomy, Nicholas Copernicus
University\\
$^4$Departament de F\'{\i}sica Aplicada, Universitat d'Alacant}

\date{\today}% It is always \today, today,
             % but any date may be explicitly specified

\begin{abstract}
  Qualitatively new aspects of the (linear and non-linear) stability
of sheared relativistic (slab) jets are analyzed. The linear
problem has been solved for a wide range of jet models well inside
the ultrarelativistic domain (flow Lorentz factors up to 20;
specific internal energies $\approx 60c^2$).  As a distinct
feature of our work, we have combined the analytical linear
approach with high-resolution relativistic hydrodynamical
simulations, which has allowed us i) to identify, in the linear
regime, resonant modes specific to the relativistic shear layer
ii) to confirm the result of the linear analysis with numerical
simulations and, iii) more interestingly, to follow the
instability development through the non-linear regime. We find
that very high-order reflection modes with dominant growth rates
can modify the global, long-term stability of the relativistic
flow. We discuss the dependence of these resonant modes on the jet
flow Lorentz factor and specific internal energy, and on the shear
layer thickness. The results could have potential applications in
the field of extragalactic relativistic jets.
% Valid PACS numbers may be entered using the \verb+\pacs{#1}+ command.
\end{abstract}

% PACS, the Physics and Astronomy %
%                             Classification Scheme.
\maketitle

\section{Introduction}

  The Kelvin-Helmholtz (KH) instability (in the simplest case, that of
a tangential discontinuity of velocity at the interface of
parallel flows) is one of the classical instabilities in fluid
dynamics. Linear perturbation analysis of KH instability has been
presented for many situations including incompressible and
compressible fluids, surface tension, finite shear layers, and
magnetized fluids \cite{Ch61}.

  The linear analysis of the KH instability for fluids in relativistic
relative motion (infinite, single vortex sheet approximation) was
developed in the seventies in the context of the stability of jets
in extended extragalactic radio sources \cite{TS76}. The main
conclusion of these studies was the reduction of the maximum
growth rate for increasing relative Lorentz factor flows and
decreasing specific internal energies (or sound speeds). The
general dispersion relation for relativistic cylindrical jets was
obtained and solved for a range of parameter combinations of
astrophysical interest \cite{FT78,Ha79}. Some approximate
analytical expressions were derived \cite{Ha87a}. General
numerical solutions of the dispersion relation were analyzed
\cite{Ha87b} and the results were applied for the first time to
the interpretation of the morphology of jets in extended radio
sources and the motion of radio components in the inner part of
these objects. Stability analysis (both in non-relativistic and
relativistic regimes) at KH instability has been used to interpret
many phenomena observed in astrophysical jets such as
quasi-periodic wiggles and knots, filaments, limb brightening and
jet disruption \cite{comment2,comment3}. More recently, KH linear
stability analysis applied to very high resolution observations
has addressed to probe the physical parameters in these sources
\cite{LZ01}.

  A general treatment of the KH instability with compressible shear
layers in the case of infinite plane boundary (non-relativistic)
problems was proposed \cite{BD75}. The study on the effects of
shear layers was extended to the case of infinite slab jets
\cite{FM82}, concentrating on the wave number range $0.1/R_j \leq
k \leq 10/R_j$ ($R_j$ is the jet radius) for ordinary ($n_x = 0$)
and the first reflection ($n_x = 1,2,3$) symmetric and
antisymmetric modes ($n_x$ represents the number of nodes across
the planar jet).

  An attempt to investigate the growth of the KH instability in some
particular class of cylindrical relativistic sheared jets was
pursued \cite{Bi91}. However, it was limited to the ordinary ($n_r
= 0$) and first two reflection modes ($n_r = 1,2$), and the domain
of jet parameters considered involved only marginally relativistic
flows (beam flow velocities $\leq 0.1 c$; $c$ is the speed of
light) and non-relativistic (jet, ambient) sound speeds ($\leq
0.01c$). Other approaches to the linear analysis of the stability
of relativistic stratified jets \cite{HS96} and sheared,
ultrarelativistic jets \cite{Ur02} have also been performed. In
the latter reference, the author has derived approximated formulae
for instability modes excited in the shear layer.

  In this paper, we report about qualitatively new aspects of the
stability of sheared relativistic (slab) jets in linear and
non-linear regimes.  We have considered a wide range of
jet/ambient parameters reaching well inside the ultrarelativistic
domain (jet flow Lorentz factors up to 20; jet specific internal
energies $\approx 60c^2$). Instead of focusing on the
stabilization effect of the shear layer on the ordinary modes
alone \cite{Bi91}, we have also studied the properties of very
high-order ($n_x \agt 20$) reflection modes which have the largest
growth rates and then dominate the global stability properties of
the flow. Finally, we have combined the analytical linear approach
with high resolution relativistic hydrodynamical simulations which
have allowed us i) to confirm the results obtained with the linear
analysis and, ii) to follow the instability development through
the non-linear regime. Our selection of the two-dimensional slab
geometry for our work responds to several reasons: i) the
possibility of using larger resolutions in two dimensional
simulations, compared to fully three dimensional simulations, ii)
the fact that slab jets allow for the study of symmetric and
antisymmetric modes, contrary to cylindric geometry that only
allows for symmetric structures, iii) it is easier to solve the
linear problem equation and to interpret results from the
numerical simulations in this case, so we can gain deep knowledge
on the physics of instabilities before studying more complex
(including three dimensions, magnetic fields...) problems. Several
recent works have combined linear analysis and hydrodynamical
simulations in connection with several astrophysical scenarios
(i.e., relativistic jets \cite{comment4a} and {\it gamma-ray
bursts} \cite{comment4b}), the relativistic nature of the jet
parameters considered (that includes the ultrarelativistic limit),
the modes explored (very high-order reflection modes), and the
complementarity of linear analysis and non-linear high-resolution
simulations make the present work unique. The results of the
numerical simulations in the nonlinear regime are presented
elsewhere \cite{Pe05}. The results shown in this paper concerning
the stability of relativistic sheared flows could be of potential
interest in the field of extragalactic relativistic jets.

\section{Instabilities in sheared relativistic jets. Linear analysis}

  We start with the equations governing the evolution of a slab
relativistic perfect-fluid jet for which the energy-momentum tensor
can be written as
\begin{equation}
T^{\mu\nu} = (\rho_e + P) u^\mu u^\nu + P
\eta^{\mu\nu}
\end{equation}
(units have been used so that $c=1$; Greek indices, $\mu$, $\nu$, run
from 0 to 3), where $\rho_e$ is the energy density, $P$ the pressure
and $u^\nu$ the fluid four-velocity.  The tensor $\eta^{\mu\nu}$ is
the metric tensor describing the geometry of the fixed, flat
space-time where the fluid evolves. In the following we will use
$u^\mu = \gamma(1,\vec{v})$, $\gamma$ being the Lorentz factor,
$\gamma = 1/\sqrt{1-v^2}$.

  The initial equilibrium configuration is that of a steady slab jet
in Cartesian coordinates flowing along the $z$-coordinate,
surrounded by a denser and colder ambient medium. A
single-component ideal gas equation of state with adiabatic
exponent $\Gamma= 4/3$ has been used to describe both jet and
ambient media. Both media are in pressure equilibrium and are
separated by a smooth shear layer of the form \cite{FM82}
\begin{equation} \label{eq:sh}
a(x) = a_\infty + (a_0-a_\infty)/\cosh (x^m),
\end{equation}
where $a(x)$ is the profiled quantity ($v_z$ and $\rho$, the rest mass
density) and $a_0$ and $a_\infty$, its value at the jet symmetry plane
(at $x=0$) and at $x \rightarrow \infty$, respectively. The integer
$m$ controls the shear layer steepness. In the limit $m \rightarrow
\infty$ the configuration tends to the vortex-sheet case.

  We now introduce an adiabatic perturbation of the form $\propto g(x)\exp
(i(k_z z - \omega t))$ in the flow equations, $\omega$ and $k_{z}$
being the frequency and wave number of the perturbation along the
jet flow. We shall follow the {\it temporal approach}, in which
perturbations grow in time having real wave numbers and complex
frequencies (the imaginary part being the {\it growth rate}). The
number of nodes across the planar jet, $n_x$, distinguishes
between ordinary modes (corresponding to $n_x = 0$) and reflection
modes ($n_x > 0$). By linearizing the equations and eliminating
the perturbations of rest mass density and flow velocity, a second
order ordinary differential equation for the pressure
perturbation, $P_1$, is obtained \cite{Bi84}
\begin{eqnarray}
  P_1^{\prime\prime} + \left(\frac{2\gamma_0^2v_{0z}^\prime (k_z - \omega
  v_{0z})}{\omega -v_{0z}k_z} - \frac{\rho_{e,0}^\prime}{\rho_{e,0} +
  P_0}\right)P_1^\prime  + & & \label{radial-eq}\\
  \gamma_0^2\left(\frac{(\omega  -v_{0z}k_z)^2}{c_{s,0}^2} - (k_z - \omega
  v_{0z})^2\right)P_1 & = & 0 \nonumber
\end{eqnarray}
where $\rho_{e,0}$ is the energy-density of the unperturbed model,
$P_0$ the pressure, $v_{0z}$ the three-velocity component, $\gamma_0 =
1/\sqrt{1-v_{0z}^2}$ is the Lorentz factor and $c_{s,0}$ is the
relativistic sound speed. The prime denotes the $x$-derivative. Unlike
the vortex sheet case, in the case of a continuous velocity profile, a
dispersion relation can not be written explicitly.
The equation (\ref{radial-eq}) is integrated from the jet axis, where boundary
conditions on the amplitude of pressure perturbation and its first derivative
are imposed

\begin{eqnarray}\label{eq:bcs1}
P_1(x = 0) = 1, & P_1^{\prime}(x = 0) = 0 & \mbox{(sym. modes)},
\\ \nonumber
P_1(x = 0) = 0, & P_1^{\prime}(x = 0) = 1  &
\mbox{(antisym. modes)}.
\end{eqnarray}
Solutions satisfying the Sommerfeld radiation conditions (no
incoming waves from infinity and wave amplitudes decaying towards
infinity) are found with the aid of the method proposed in
Ref.\cite{RL84}, based on the shooting method \cite{Press97}.

%%%%%%%%%%%%%%%%%%%%%%%%%%%%%%%%%%%%%%%%%%%%%%%%%%%%%%%%%%%%%%%%%%%%%
%
%Fig.1

\begin{figure*}
\includegraphics[width=0.48\textwidth]{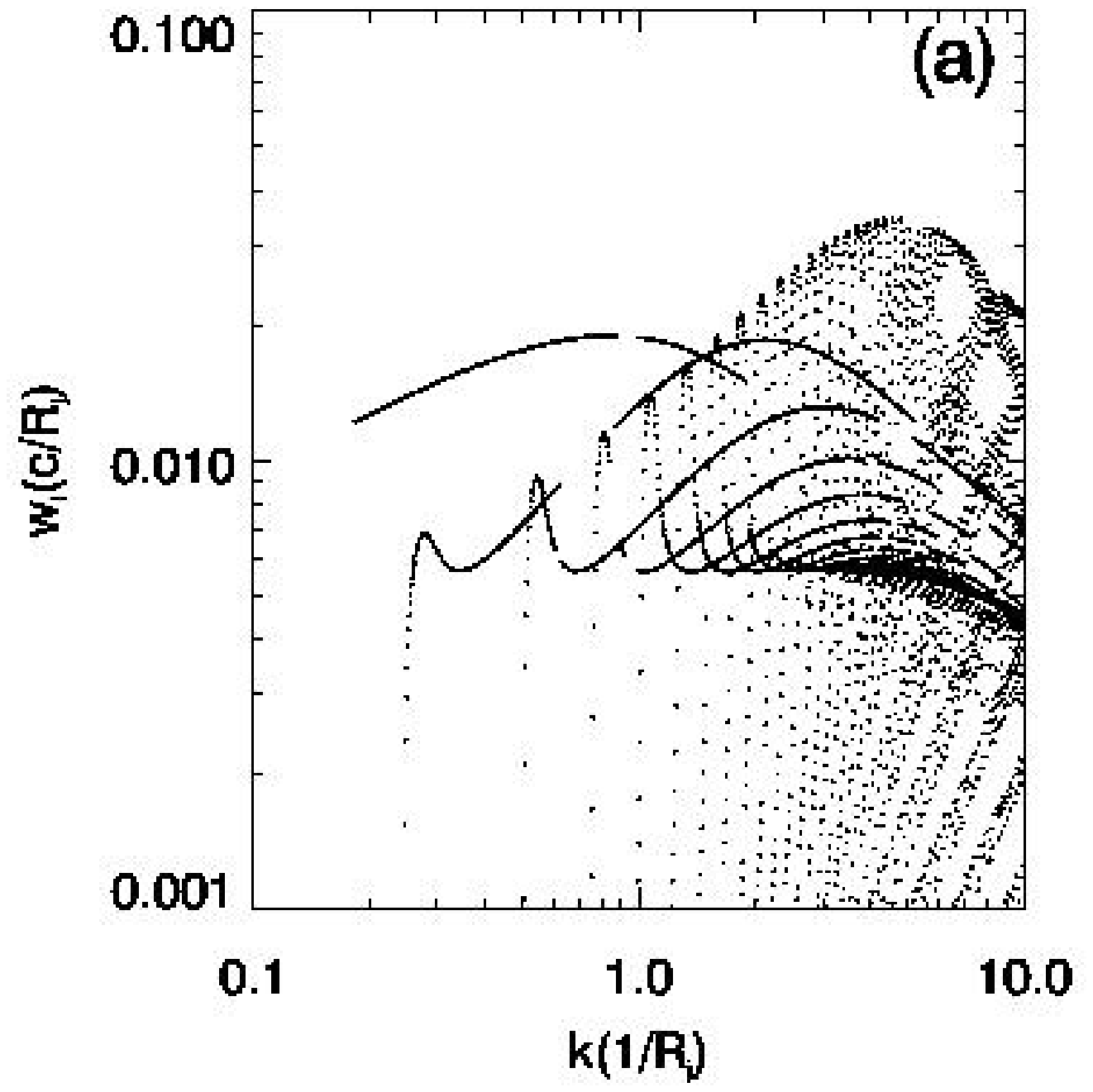}
\includegraphics[width=0.48\textwidth]{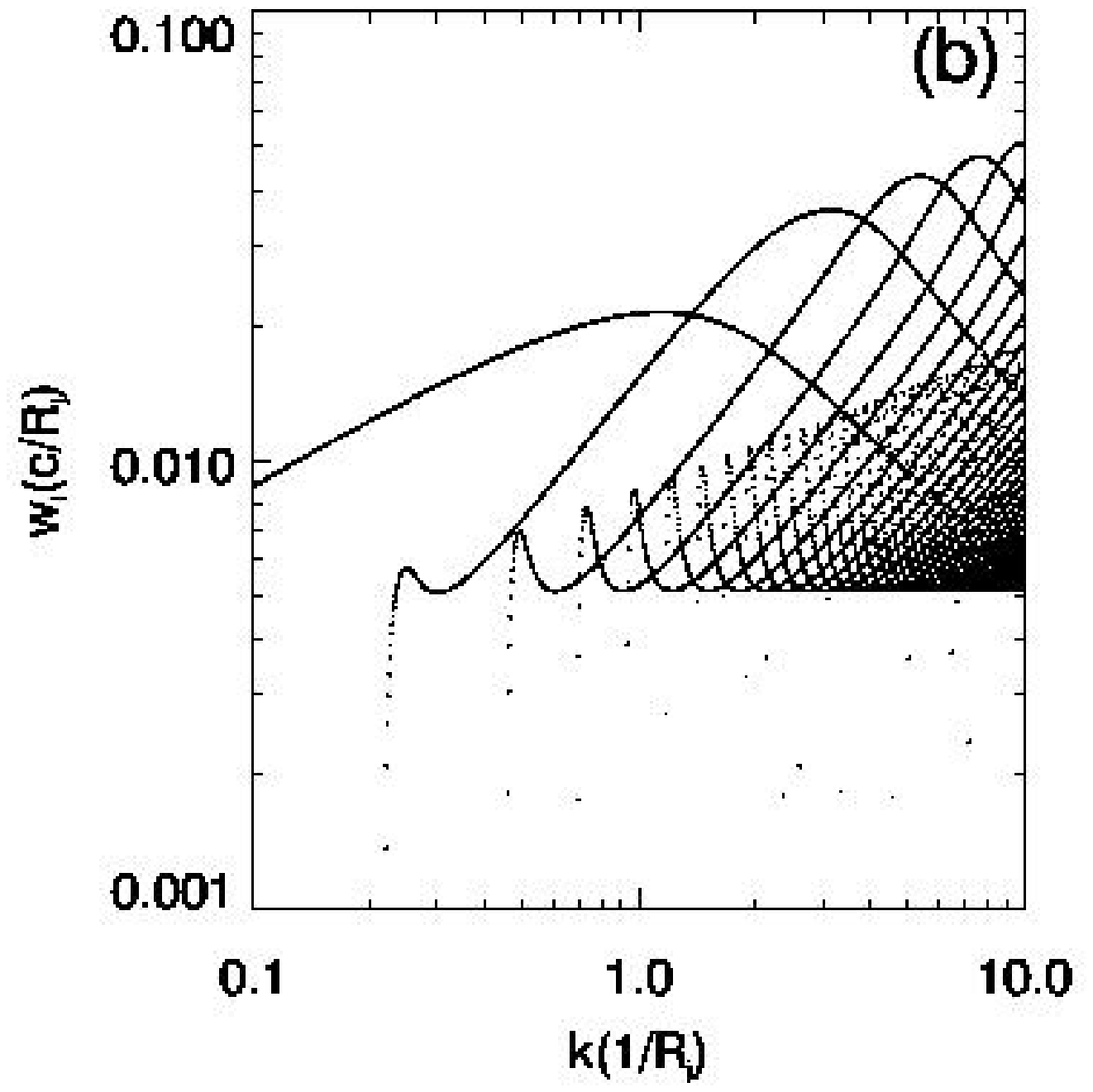}

\caption{\label{fig:f1} Growth rate vs. longitudinal wave number
for Model D20, using a shear layer with $m = 25$ in
Eq.~(\ref{eq:sh}) (panel a) and vortex sheet (panel b) for the
fundamental and a series of reflection, antisymmetric modes
including the one with the absolute maximum in the growth rate.
Main differences are the overall decrease of growth rates in the
sheared case, and the appearance in this case of sharp resonances
at the small wave number limit for each high-order reflection mode
with the largest growth rates for a given mode.}

\end{figure*}
%
%%%%%%%%%%%%%%%%%%%%%%%%%%%%%%%%%%%%%%%%%%%%%%%%%%%%%%%%%%%%%%%%%%%%%

  We have solved the linear problem for more than 20 models with
different specific internal energies of the jet, Lorentz factors
and shear layer widths, fixing jet/ambient rest-mass density
contrast ($ = 0.1$). We used $m = 8, 25, 2000$ (shear layer width,
$d \approx 0.6, 0.177, 5\, 10^{-3} R_j$) and vortex sheet for jets
having specific internal energies $\varepsilon_j = 0.4c^2$ (models
B) and $60c^2$ (models D) and Lorentz factors $\gamma_j = 5$ (B05,
D05) and $20$ (B20, D20). Solutions with $m = 2000$ were
considered in order to test convergence to vortex sheet in the
case of narrow shear layers, with positive results. Also, fixing
the width of the shear layer by setting $m = 25$, we solved for
$\varepsilon_j = 0.7c^2$ (model A), along with models B and D,
using $\gamma_j = 2.5$ and $10$, in order to span a wide range of
parameters \cite{PH04}.

%%%%%%%%%%%%%%%%%%%%%%%%%%%%%%%%%%%%%%%%%%%%%%%%%%%%%%%%%%%%%%%%%%%%%%%%%%%
%
%Fig.2

\begin{figure}
\includegraphics[width=1.4\columnwidth,angle=270]{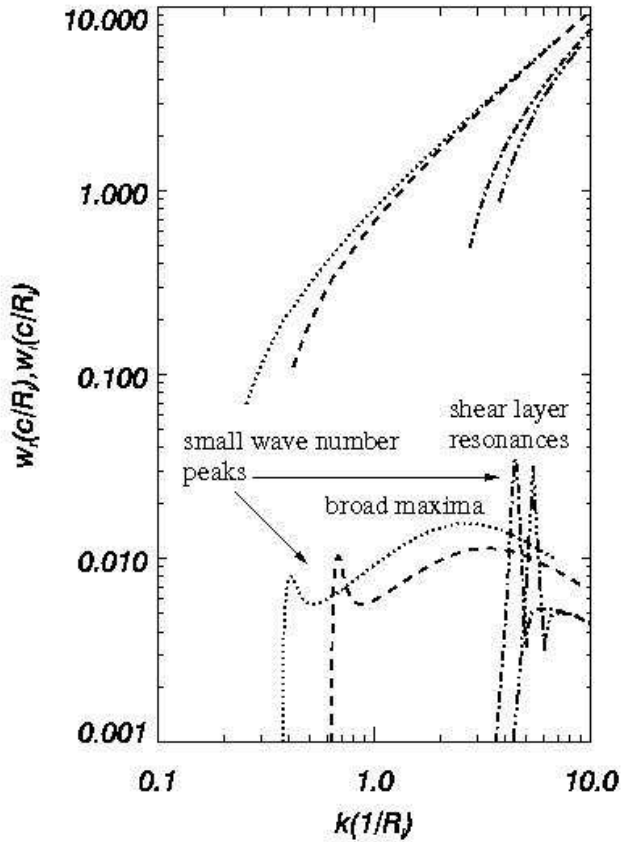}
\caption{Specific symmetric modes of Model D20. Dotted line: first
reflection mode, dashed: second reflection mode, dash-dot:
twentieth reflection mode, dash-triple dot: twenty-fifth
reflection mode. We point both the broad maxima and the small wave
number peaks present in every single mode. Small wave number peaks
of high order reflection modes show larger growth rates and are
thus defined as (shear layer) resonances.} \label{fig:f2}
\end{figure}
%
%%%%%%%%%%%%%%%%%%%%%%%%%%%%%%%%%%%%%%%%%%%%%%%%%%%%%%%%%%%%%%%%%%%%%%%%%%%%

  The effect of the shear layer on the linear stability is seen in
Fig.~\ref{fig:f1} where we show the growth rates of the fundamental
and a series of reflecting (antisymmetric) modes resulting from the
solution of the equation (\ref{radial-eq}) together with the boundary
conditions (\ref{eq:bcs1}) for Model D20. The corresponding solution
for the vortex sheet case is also shown for comparison.

%%%%%%%%%%%%%%%%%%%%%%%%%%%%%%%%%%%%%%%%%%%%%%%%%%%%%%%%%%%%%%%%%%%%%
%
%Fig.3

\begin{figure*}
\includegraphics[width=0.33\textwidth]{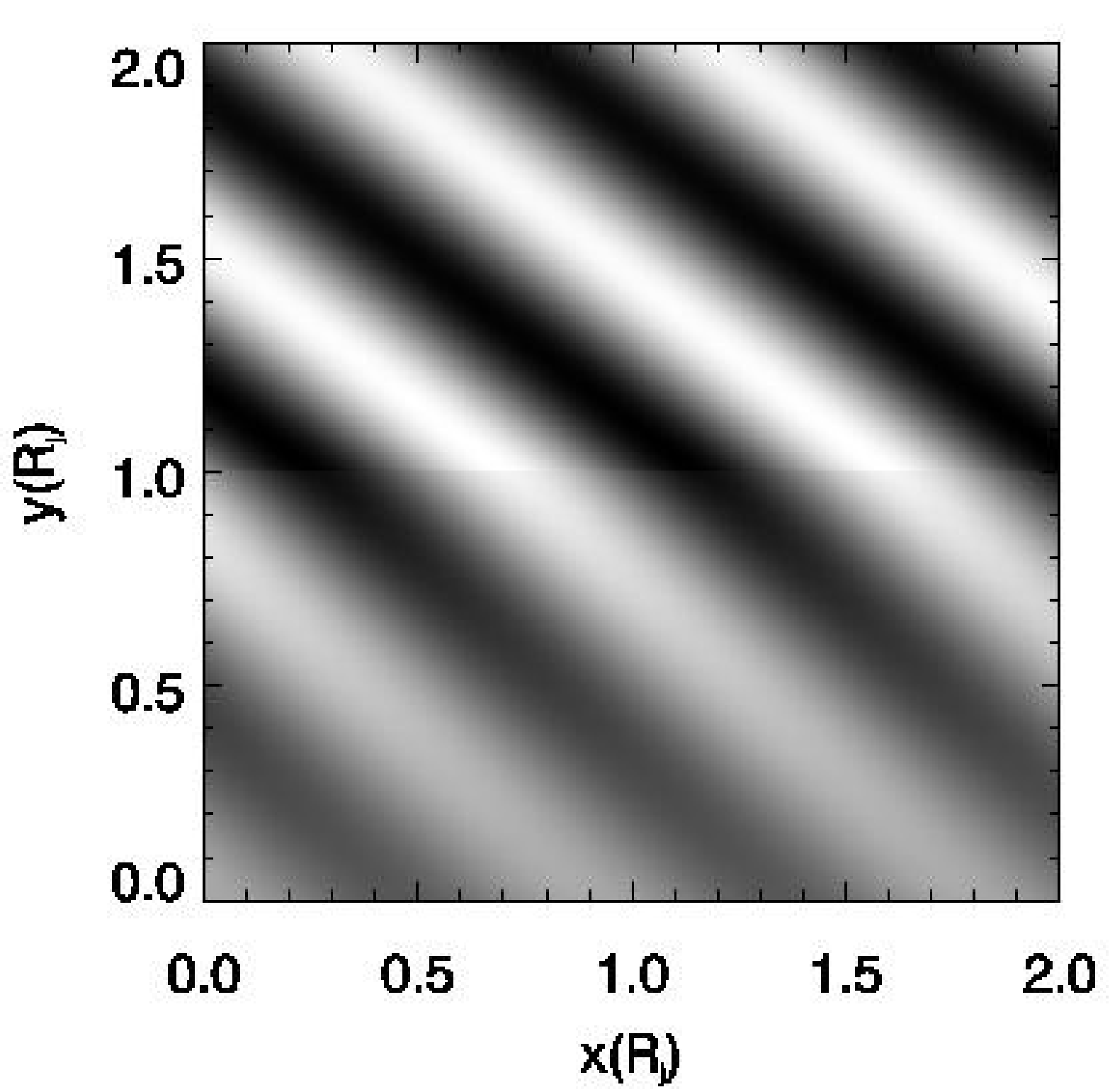}
\nolinebreak
\includegraphics[width=0.33\textwidth]{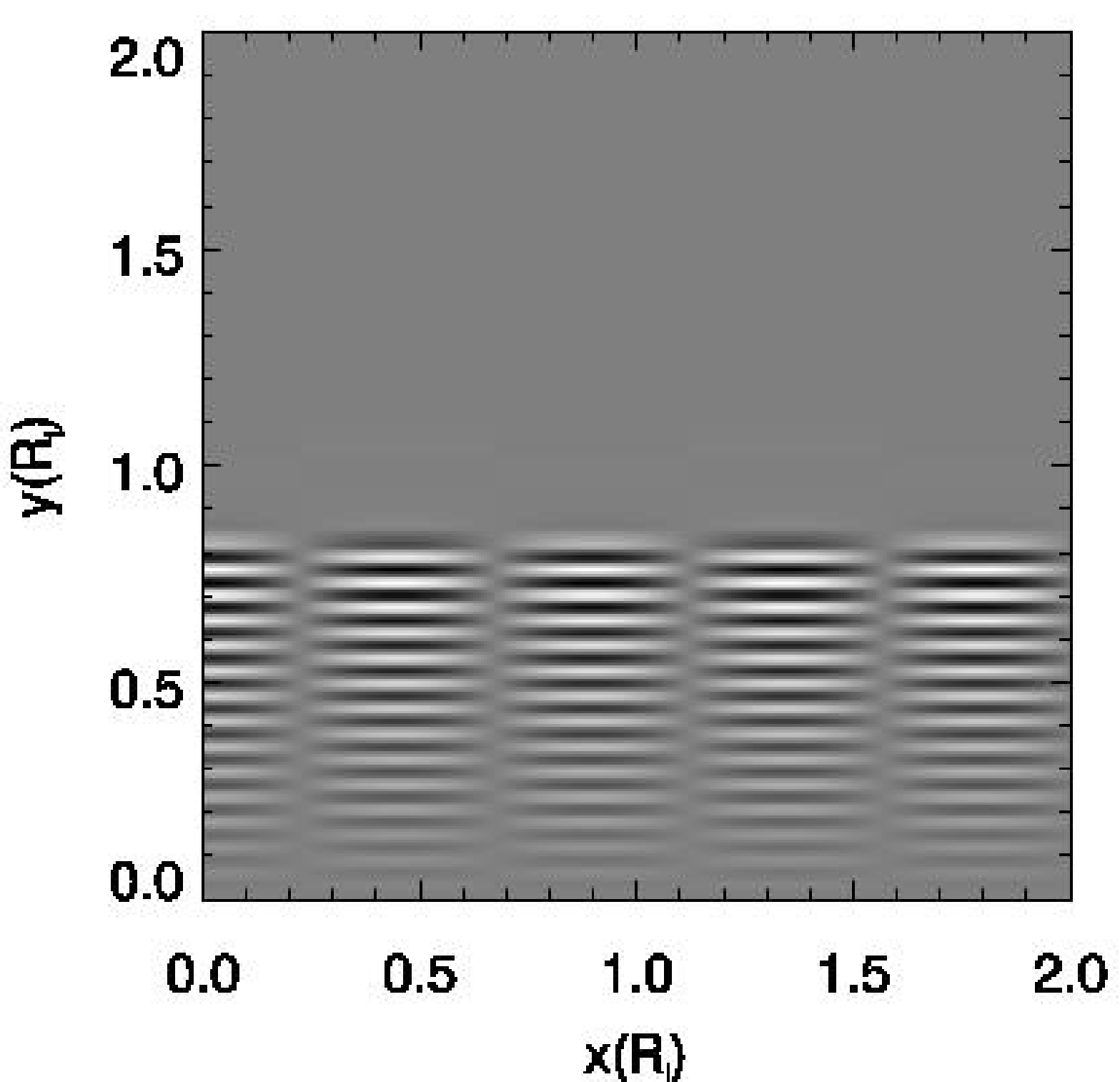}
\nolinebreak \includegraphics[width=0.33\textwidth]{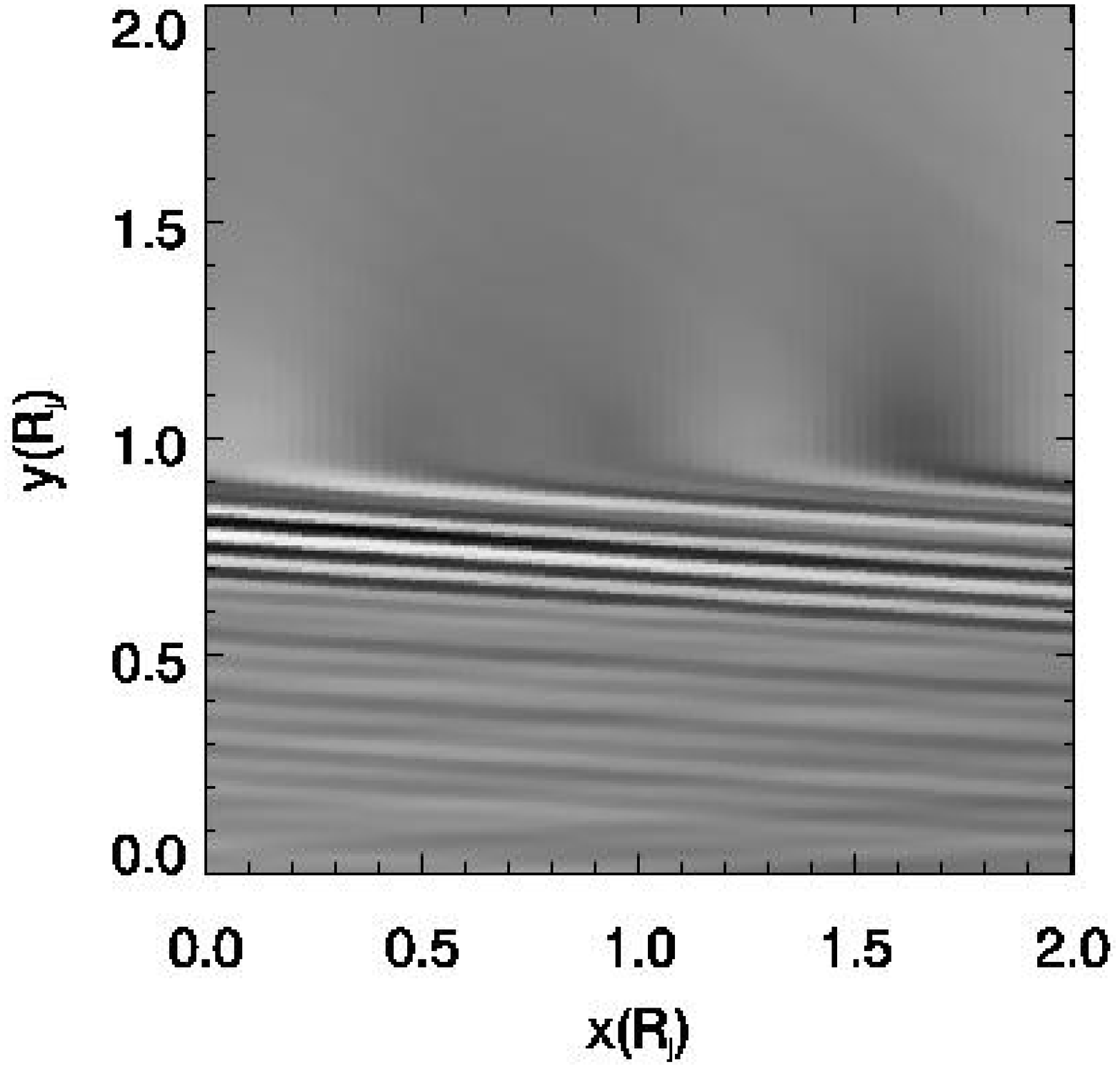}

\caption{\label{fig:f3} Two-dimensional panels of different
pressure perturbation structures for Model D20.  The gray scale
extends over the pressure variations (in arbitrary units). Lengths
are measured in (initial) jet radii, $R_j$. Flow is from left to
right and periodical. The bottom boundary corresponds to the jet
symmetry plane. Left panel: vortex sheet dominant mode (low order
reflection mode) at a given wavelength (from linear solution).
Central panel: Dominant mode (high order reflection mode) at the
same wavelength when $m = 25$ (Eq.~\ref{eq:sh}) shear layer is
included (also from linear solution). Right panel: Pressure
perturbation map from a hydrodynamical simulation in the linear
regime. The resolution used in the simulation was 256 cells/$R_j$
across the jet and 32 cells/$R_j$, along. Grid size was 6 $R_j$
transversally and 8 $R_j$ axially, with an extended, decreasing
resolution, grid in the transversal direction up to 100 $R_j$.
Periodic boundary conditions were applied at the left and right
ends of the grid, and outflow boundary conditions far from the jet
in the transversal direction.}

\end{figure*}
%
%%%%%%%%%%%%%%%%%%%%%%%%%%%%%%%%%%%%%%%%%%%%%%%%%%%%%%%%%%%%%%%%%%%%%

  We note that the reflection mode solutions of the shear problem are
more stable (i.e., the growth rates are smaller) for most wave
numbers, especially in the large wave number limit, than the
corresponding solutions in the vortex sheet case. This behaviour
was reported for the first time for the first and second
reflection modes in the non-relativistic limit \cite{FM82}. The
growth rate curves corresponding to a single $n$-th reflection
mode consists of a broad maximum at larger wave numbers and a
local peak which is placed in the small wave number limit, near
the marginal stability point of the mode. While in the
relativistic jet, vortex-sheet case the small wave number peaks
are relatively unimportant (since the maximum growth rates at
these peaks are lower than the growth rates of other unstable
modes), in the presence of the shear-layer they significantly
dominate over other modes.  Therefore we shall call these peaks
{\em the shear layer resonances} \cite{comment5}. In
Fig.~\ref{fig:f2} we show the solution for four specific symmetric
modes (two low order and two high order reflection modes) of Model
D20. Low order modes do not show strong peaks at maximum unstable
wavelengths, whereas high order reflection modes show peaks (the
so-called shear layer resonances) at this maximum wavelength and
do not present broad maxima. The dependence of the properties of
the growth rates associated to the shear layer resonances on the
jet specific internal energy, jet Lorentz factor and shear layer
parameter $m$ can be summarized as follows: i) An increase of the
jet Lorentz factor enhances the dominance of resonant modes with
respect to ordinary and low order reflection modes; ii) a decrease
in the specific internal energy of the jet causes resonances to
appear at longer wavelengths; iii) a widening of the shear layer
reduces the growth rates and the dominance of the shear-layer
resonances, suggesting that there is an optimal width of the shear
layer that maximizes the effect, for a given set of jet
parameters; iv) as the shear layer widens, the largest growth rate
of resonant modes moves towards smaller wave numbers and lower
order reflection modes; v) modes with wave number larger than some
limiting value that decreases with the shear layer width are
damped significantly, consistently with previous non-relativistic
results \cite{FM82}.

%%%%%%%%%%%%%%%%%%%%%%%%%%%%%%%%%%%%%%%%%%%%%%%%%%%%%%%%%%%%%%%%%%%%%
%
%Fig.4
\begin{figure*}
\includegraphics[width=0.3\textwidth]{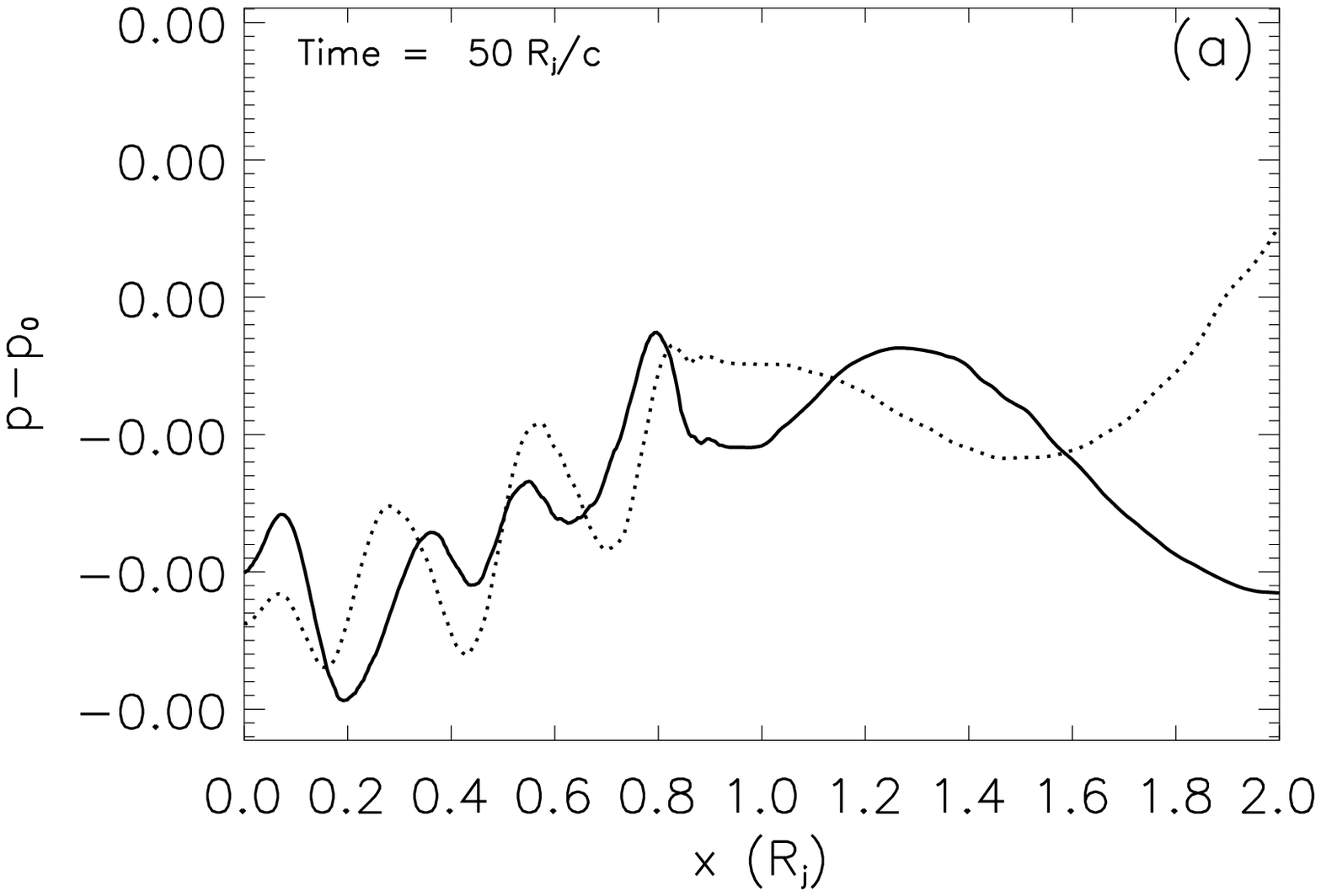}
\includegraphics[width=0.3\textwidth]{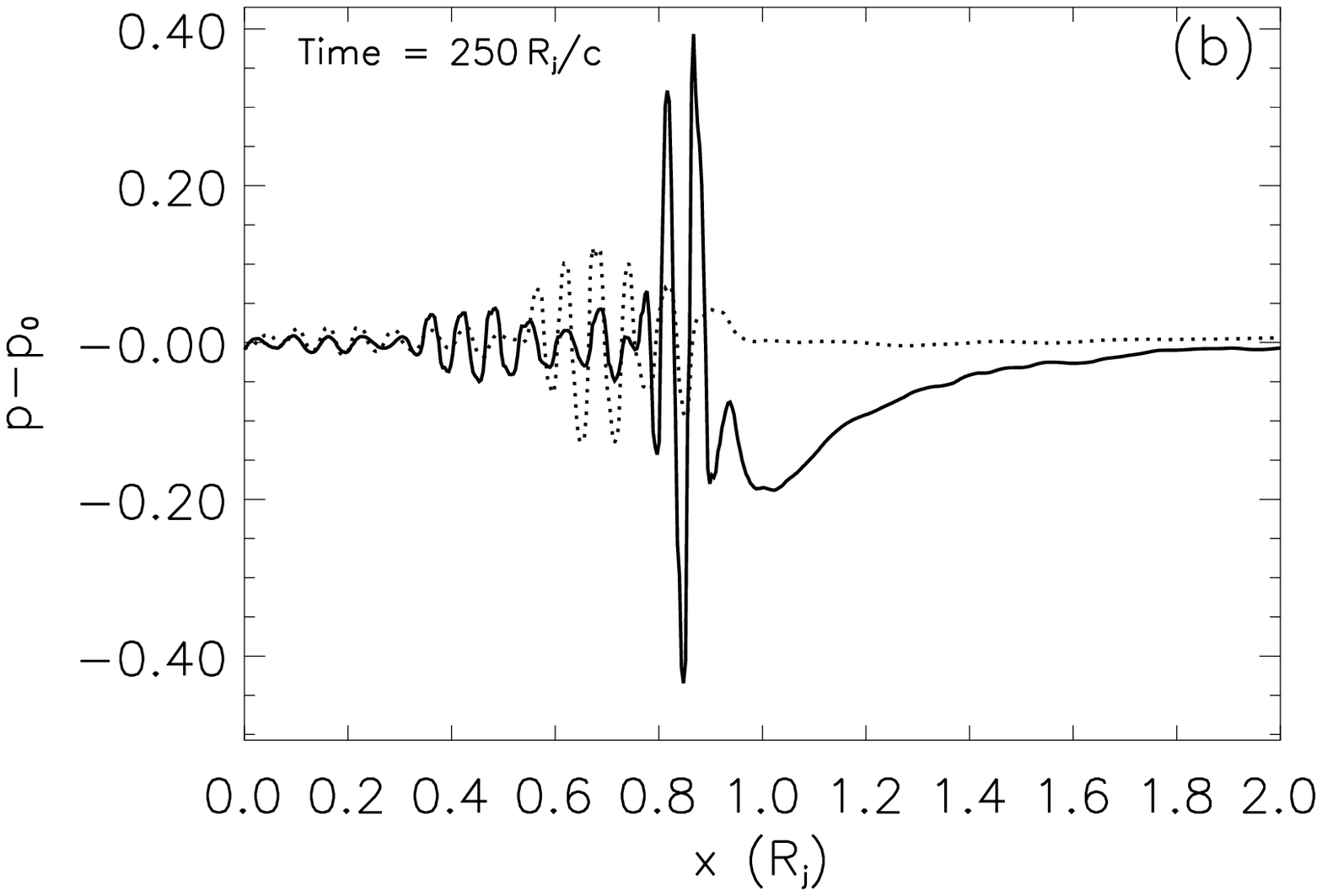}
\includegraphics[width=0.3\textwidth]{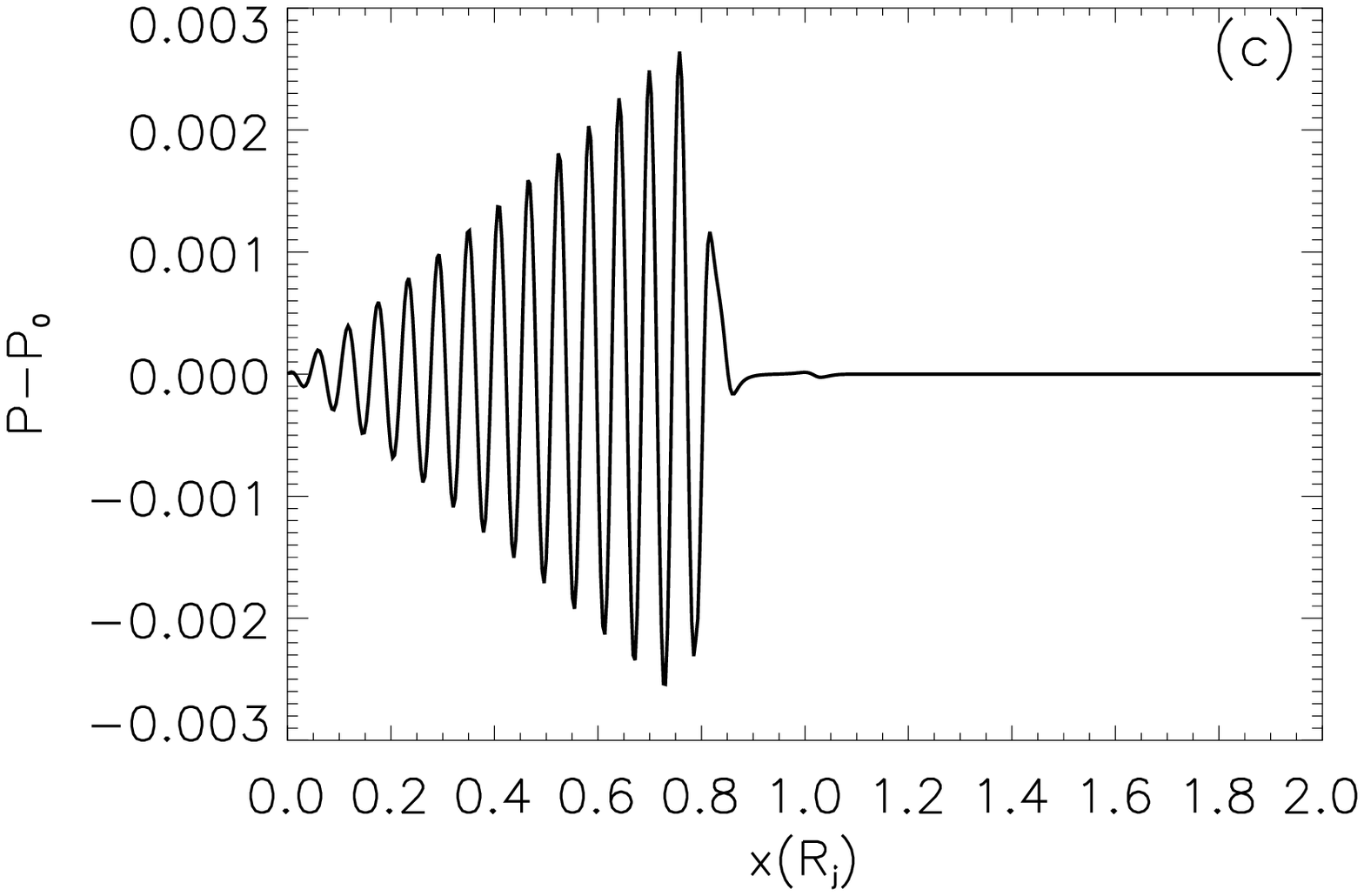}

\caption{\label{fig:f4} Radial plots of pressure perturbation
($P-P_0$, with $P_0=2.0\, \rho_{ext}\,c^2$) at two different times
in the simulation for Model D20 (see the caption of
Fig.~\ref{fig:f3} for details) and a theoretical representation of
the transversal structure of the fastest growing resonant mode, at
the wavelength observed in the simulation, in arbitrary units
(panel c). The left panel (a) shows the perturbation in a moment
when the resonant modes have not appeared yet, and the central
panel (b) shows a moment when the resonant modes dominate the
linear regime. The solid line stands for pressure perturbation at
$z=0\,R_j$ and dotted line stands for the pressure perturbation at
half grid $z=4\,R_j$.}
\end{figure*}
%
%%%%%%%%%%%%%%%%%%%%%%%%%%%%%%%%%%%%%%%%%%%%%%%%%%%%%%%%%%%%%%%%%%%%%
  The shear layer resonances correspond to very distinct spatial
structures of eigenmodes. In Fig.~\ref{fig:f3}, we show maps of
different structures generated in a jet by pressure perturbation,
depending on the excited KH mode, as derived by theory and
simulations. The structure of maximally unstable eigenmodes in the
vortex sheet case and non-resonant modes in the sheared case (left
panel of Fig.~\ref{fig:f3}) represents a superposition of oblique
sound waves in both the jet interior and the ambient medium.
Contrarily, in the shear layer case (central panel of
Fig.~\ref{fig:f3}), the most unstable resonant modes have a very
large transversal wave number (the transversal wavelength is
comparable to the width of the shear layer) in the jet interior
and they are strongly damped in amplitude in the ambient medium.
In order to demonstrate the relevance of the resonant modes in the
evolution of the flow, we display in the right panel of
Fig.~\ref{fig:f3} an analogous pressure map resulting from a
numerical hydrodynamical simulation \cite{MM97}. In this
simulation an equilibrium jet corresponding to Model D20 with $m =
25$ (the value of $m$ is 25 for all the numerical simulations
presented here, unless explicitly indicated) has been perturbed
with a superposition of small amplitude sinusoidal perturbations.
The pressure snapshot displayed in the right panel of
Fig.~\ref{fig:f3} corresponds to an early stage of the evolution
in which the perturbation is still small (linear phase). The
resonant mode starts to dominate in the numerical simulation due
to its large growth rate, and its spatial structure is very
similar to that of the most unstable (resonant) eigenmode obtained
from the corresponding linear problem (central panel of
Fig.~\ref{fig:f3}).

Fig.~\ref{fig:f4} shows two radial plots of the pressure
perturbation, corresponding to Model D20 introduced in the
previous paragraph, at two different times (panels a and b) during
the linear phase, before (panel a) and after (panel b) the
resonant modes become dominant in the jet structure, emphasizing
the conclusions derived from Fig.~\ref{fig:f3}. A clear change in
the transversal structure of the perturbation is observed, where
the radial structure (small radial wavelength) of the growing
resonant mode is displayed in the right panel (c) of
Fig.~\ref{fig:f4}. In Fig.~\ref{fig:f4}c, the theoretical profile
of the fastest growing resonant mode, at the wavelength found in
the simulation, is shown. The number of zeros in the central and
right panels is the same (29 in both cases), implying a correct
identification of the mode (29th body mode). A difference in the
amplitude profile of the mode is observed between this theoretical
structure and that found in the simulations (Fig.~\ref{fig:f4}b).
This is due to a growth of the modes faster than predicted by the
theory in the shear layer, which might be caused by interactions
between waves. The modulation of amplitudes observed in
Fig.~\ref{fig:f4}b for radii $r<0.8\,R_j$, gives support to the
idea of interference between modes.

\section{Non-linear evolution}

%%%%%%%%%%%%%%%%%%%%%%%%%%%%%%%%%%%%%%%%%%%%%%%%%%%%%%%%%%%%%%%%%%%%%
%
%Fig.5

\begin{figure}[!h]
\includegraphics*[width=0.48\columnwidth]{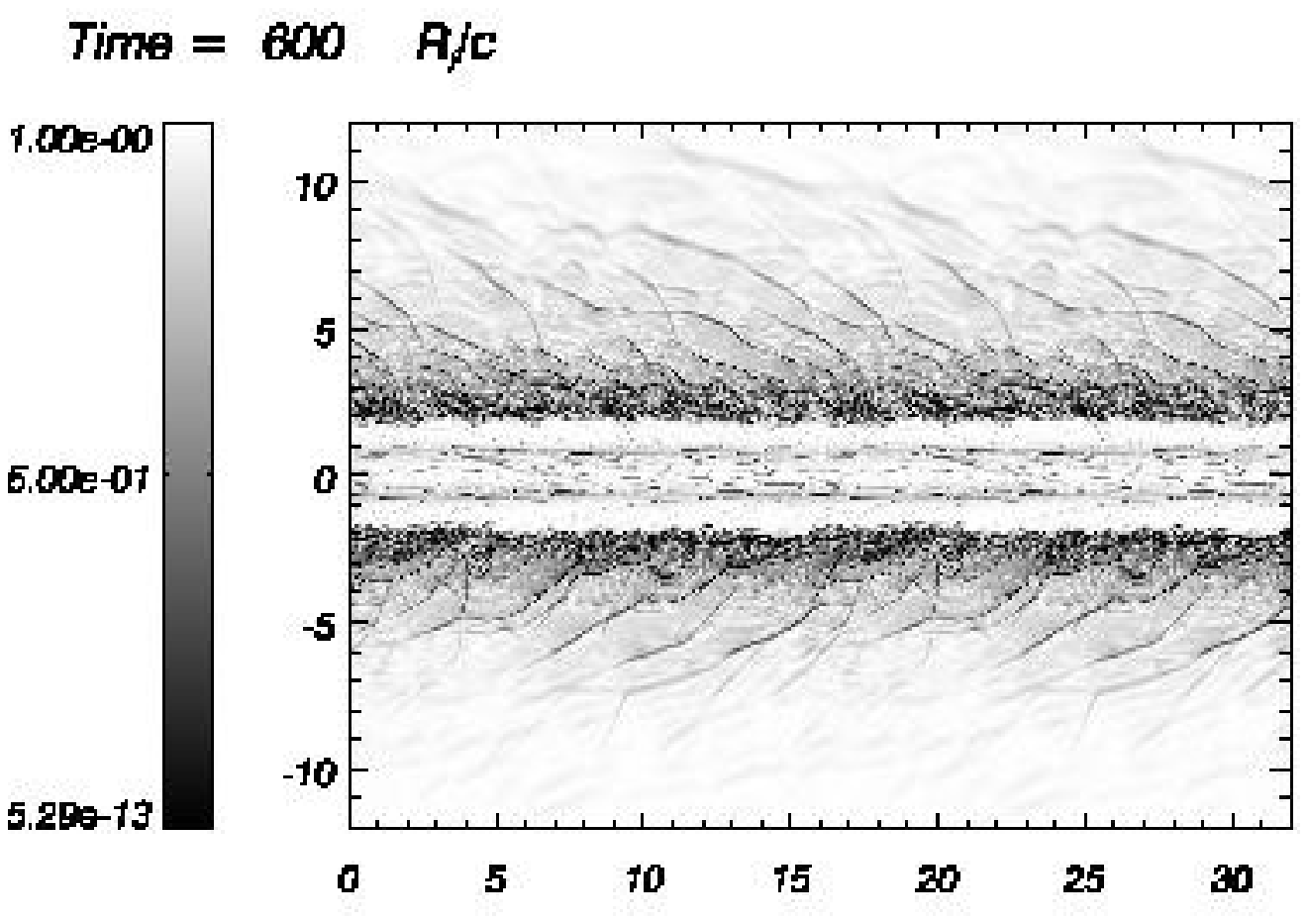}
\includegraphics*[width=0.48\columnwidth]{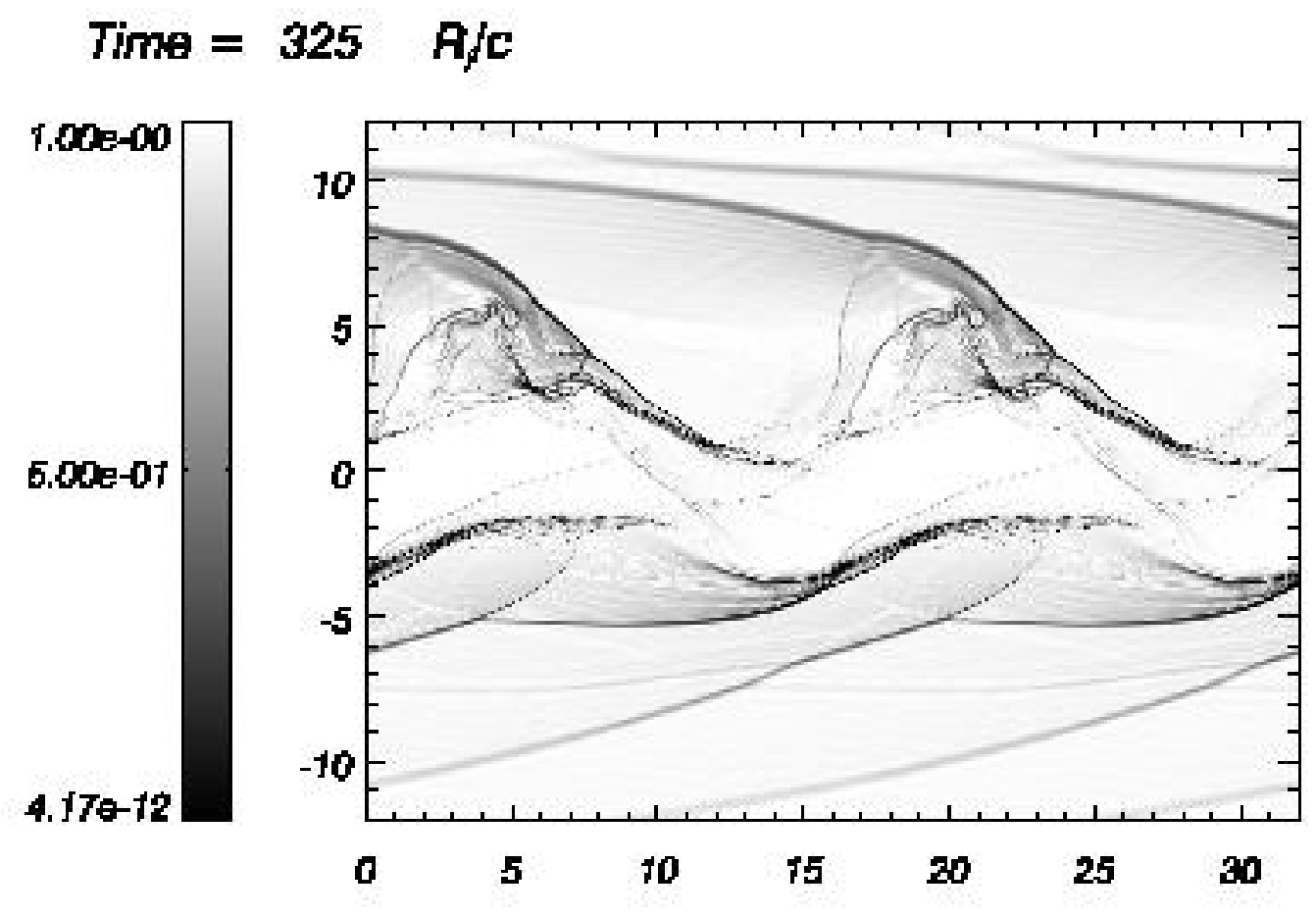}

\includegraphics*[width=0.48\columnwidth]{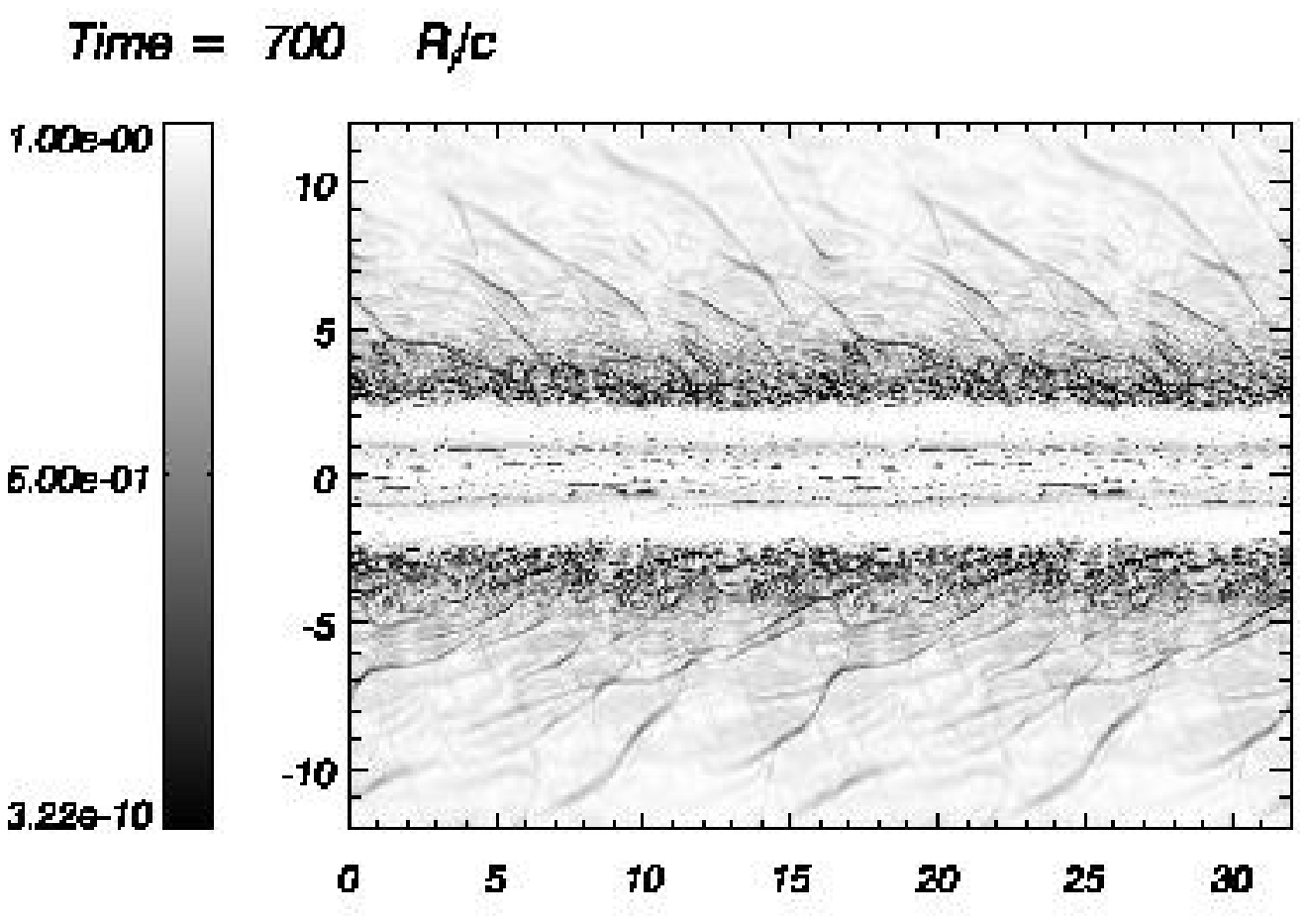}
\includegraphics*[width=0.48\columnwidth]{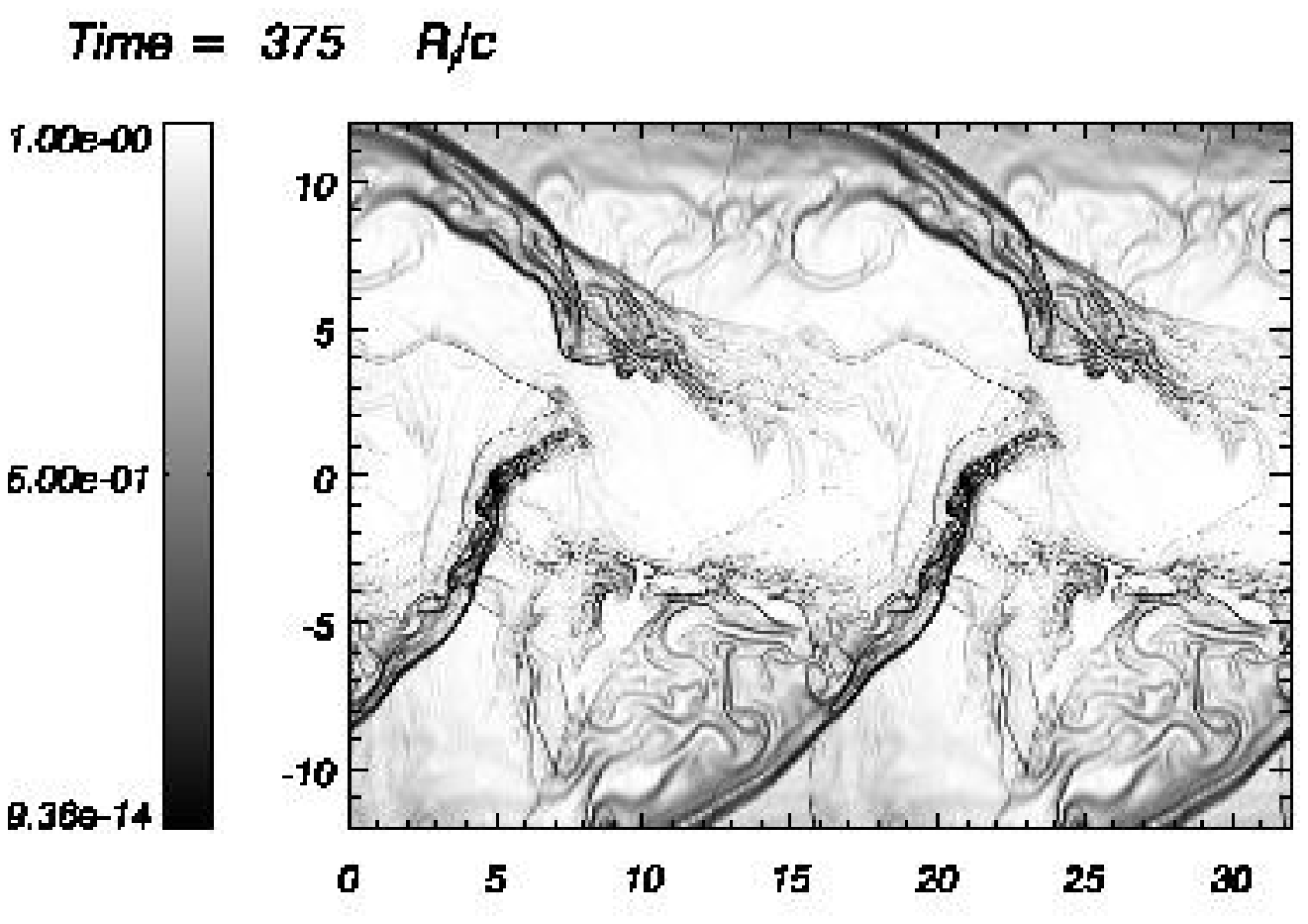}

\includegraphics*[width=0.48\columnwidth]{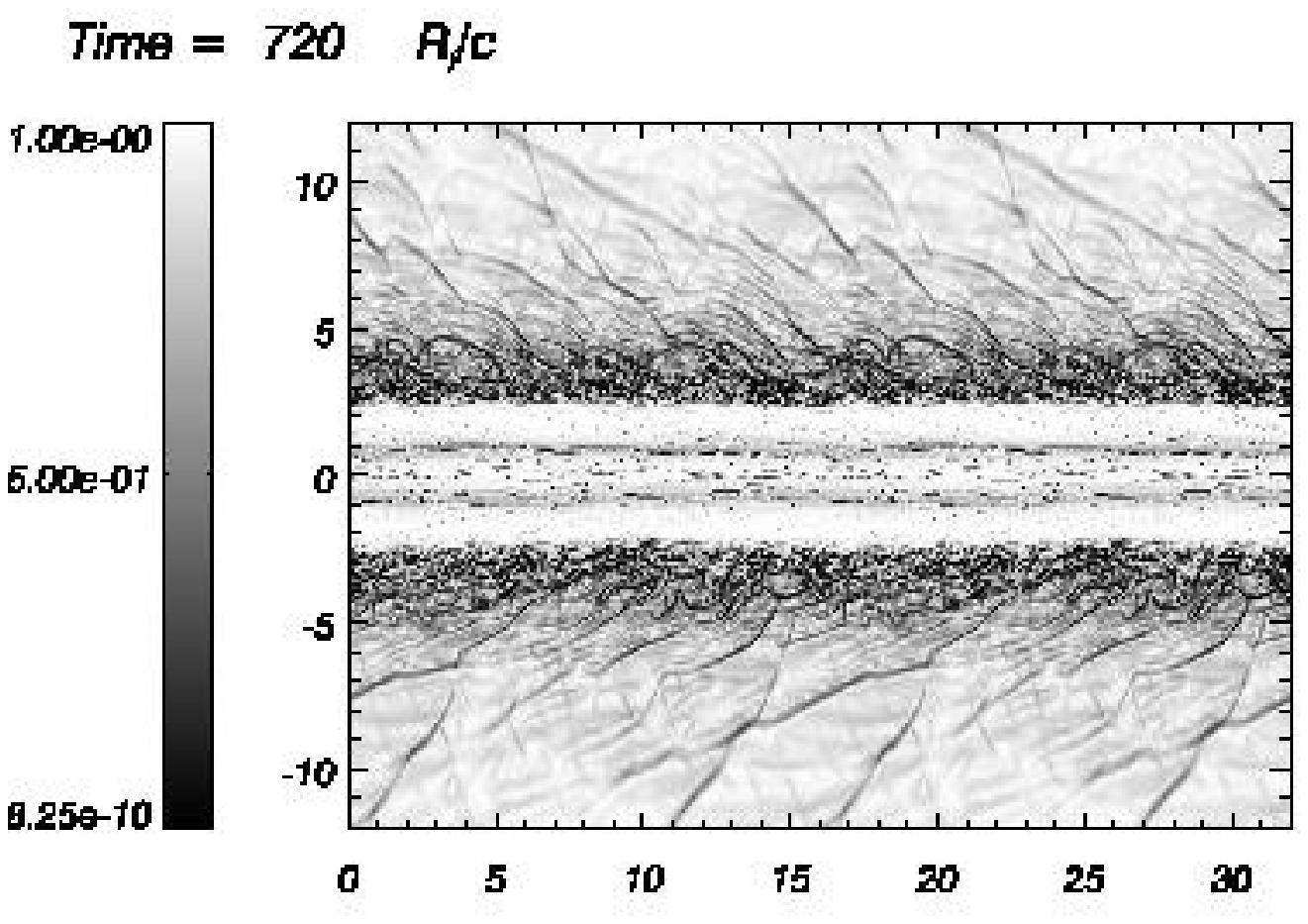}
\includegraphics*[width=0.48\columnwidth]{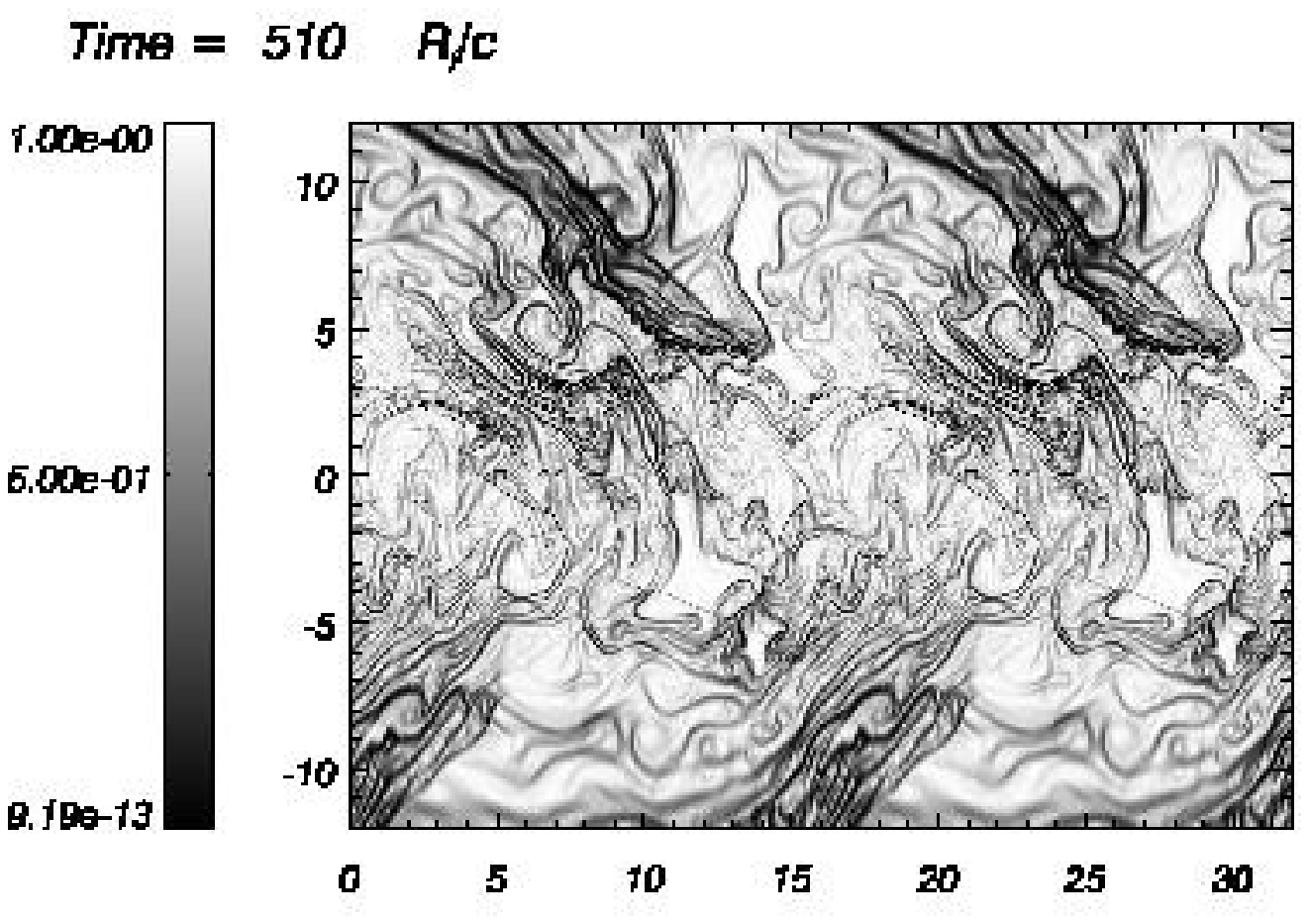}

\caption{\label{fig:f5} Schlieren plots at different times in the
non-linear regime for models B20 (left panels), at times
$t=600,\,700$ and $720\,R_j/c$ and B05 (right panels) at times
$t=325,\,375$ and $510\,R_j/c$. Shear layer resonances shield the
jet in Model B20 against disruption. Grid size was 6 $R_j$
transversally and 16 $R_j$ axially in these simulations (see the
caption of Fig.~\ref{fig:f3} for further details).}

\end{figure}
%
%%%%%%%%%%%%%%%%%%%%%%%%%%%%%%%%%%%%%%%%%%%%%%%%%%%%%%%%%%%%%%%%%%%%%

  The importance of the shear-layer resonant modes relies not only on
their dominance among solutions of the linearized problem. The
numerical simulations show that whenever these modes appear
(mostly in models with both high Lorentz factor and high
relativistic Mach number) the transition of the overall perturbed
jet structure to nonlinear regime is significantly altered. In
Fig.~\ref{fig:f5} we show how  the resonant modes affect the
non-linear evolution of instabilities in jets with larger Lorentz
factors and relativistic Mach numbers. The maps represent
schlieren plots for Model B20 ($\gamma=20$, left panels) and Model
B05 ($\gamma=5$, right panels). Model B20 shows a well collimated
jet with only small scale variations in time, due to the
development of resonant modes, whereas the jet in Model B05
undergoes strong sideways oscillations which lead to the formation
of strong oblique shocks (first panel) and the subsequent jet
disruption.

%%%%%%%%%%%%%%%%%%%%%%%%%%%%%%%%%%%%%%%%%%%%%%%%%%%%%%%%%%%%%%%%%%%%%
%
%Fig.6

\begin{figure*}
\includegraphics[width=0.9\textwidth]{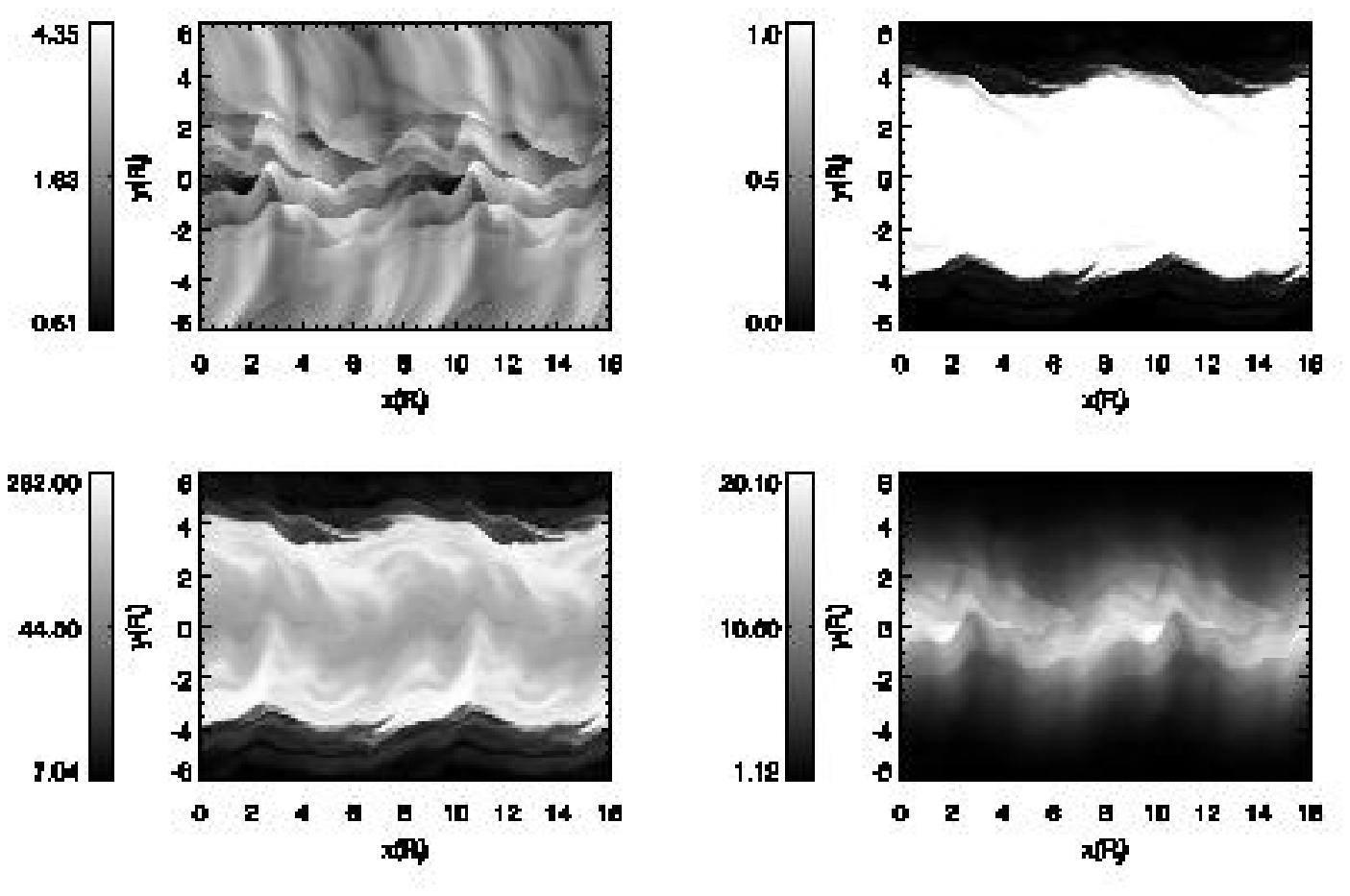}
%\vspace{6cm}
\caption{Two-dimensional panels of logarithm of pressure (top
left), tracer (top right), logarithm of specific internal energy
(bottom left) and Lorentz factor (bottom right) of Model D20 at $t
= 1000 R_j/c$ well inside the non-linear regime and once an
asymptotic quasi-steady state has been reached. Lengths are
measured in (initial) jet radii, $R_j$. Initial tracer values are
1.0 for pure jet matter and 0.0 for pure ambient matter. As seen
in the tracer panel, the final width of the jet is three times the
initial one. A thick shear layer with high specific internal
energy is observed in the bottom-left panel.} \label{fig:f6}
\end{figure*}
%
%%%%%%%%%%%%%%%%%%%%%%%%%%%%%%%%%%%%%%%%%%%%%%%%%%%%%%%%%%%%%%%%%%%%%

%%%%%%%%%%%%%%%%%%%%%%%%%%%%%%%%%%%%%%%%%%%%%%%%%%%%%%%%%%%%%%%%%%%%%
%
%Fig.7

\begin{figure*}
\includegraphics[width=0.9\textwidth]{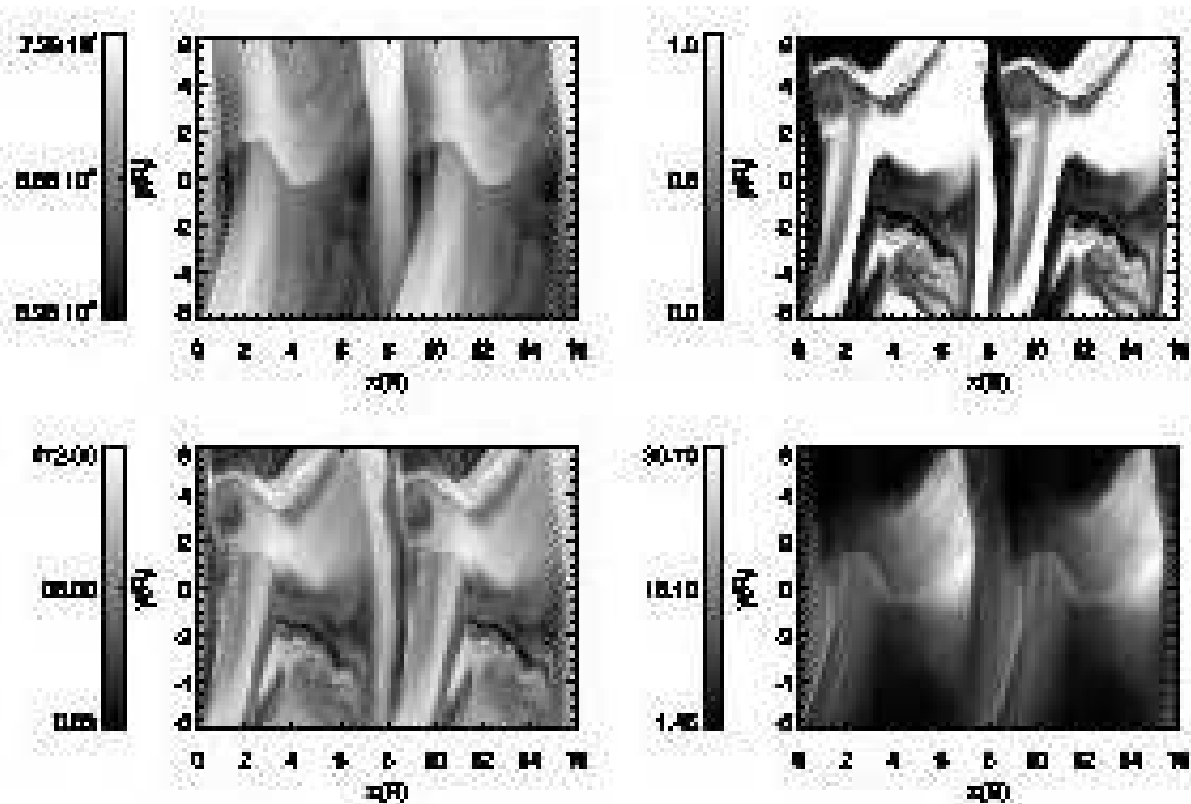}
%\vspace{6cm}
\caption{Two-dimensional panels of of pressure (top left), tracer
(top right), logarithm of specific internal energy (bottom left)
and Lorentz factor (bottom right) of Model D20, in the
vortex-sheet analytical limit, at $t = 595 R_j/c$. Compare with
Fig.~\ref{fig:f6}.} \label{fig:f7}
\end{figure*}
%
%%%%%%%%%%%%%%%%%%%%%%%%%%%%%%%%%%%%%%%%%%%%%%%%%%%%%%%%%%%%%%%%%%%%%

%%%%%%%%%%%%%%%%%%%%%%%%%%%%%%%%%%%%%%%%%%%%%%%%%%%%%%%%%%%%%%%%%%%%%
\begin{figure*}[!t]
\includegraphics[width=0.48\textwidth]{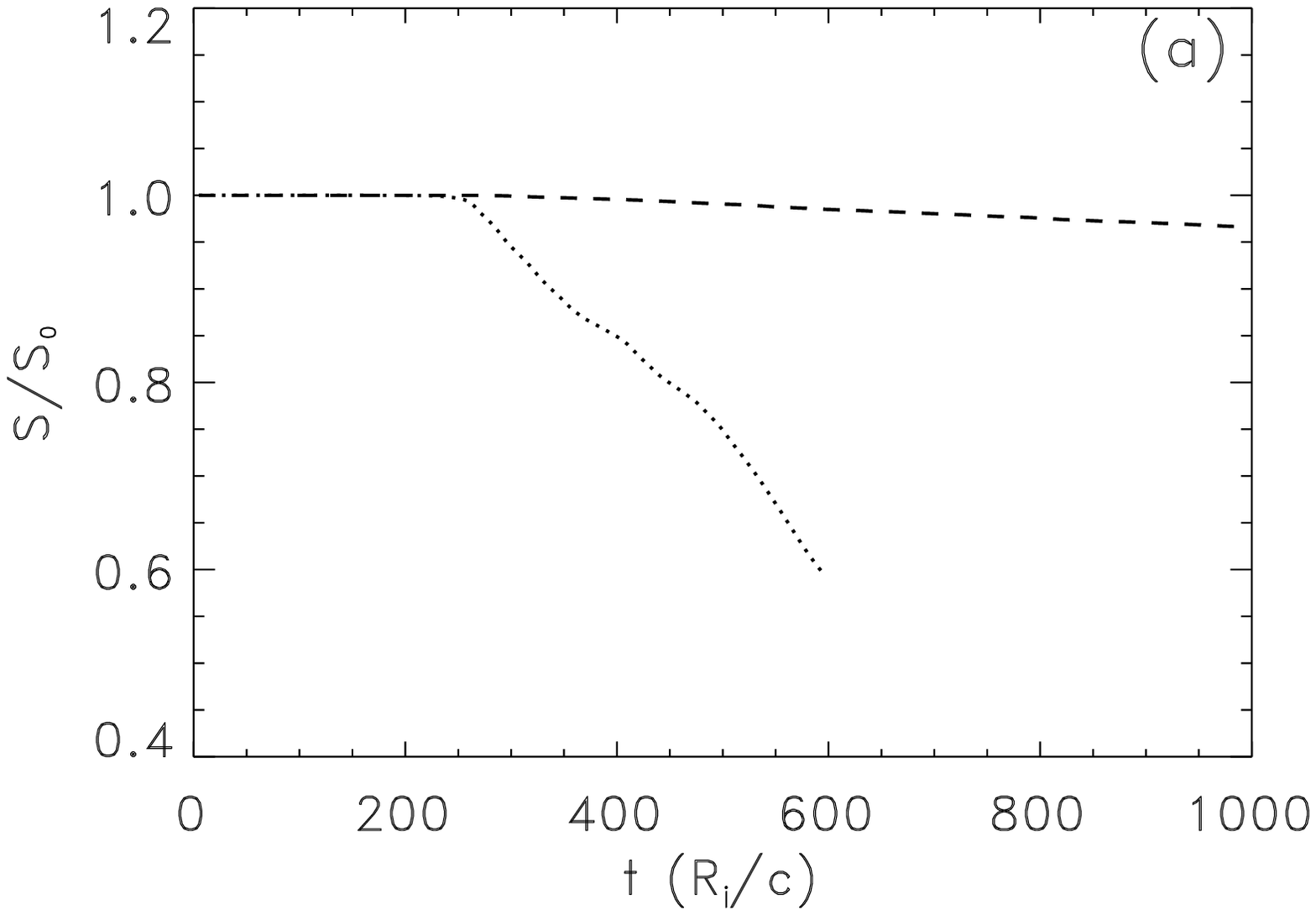}
\includegraphics[width=0.48\textwidth]{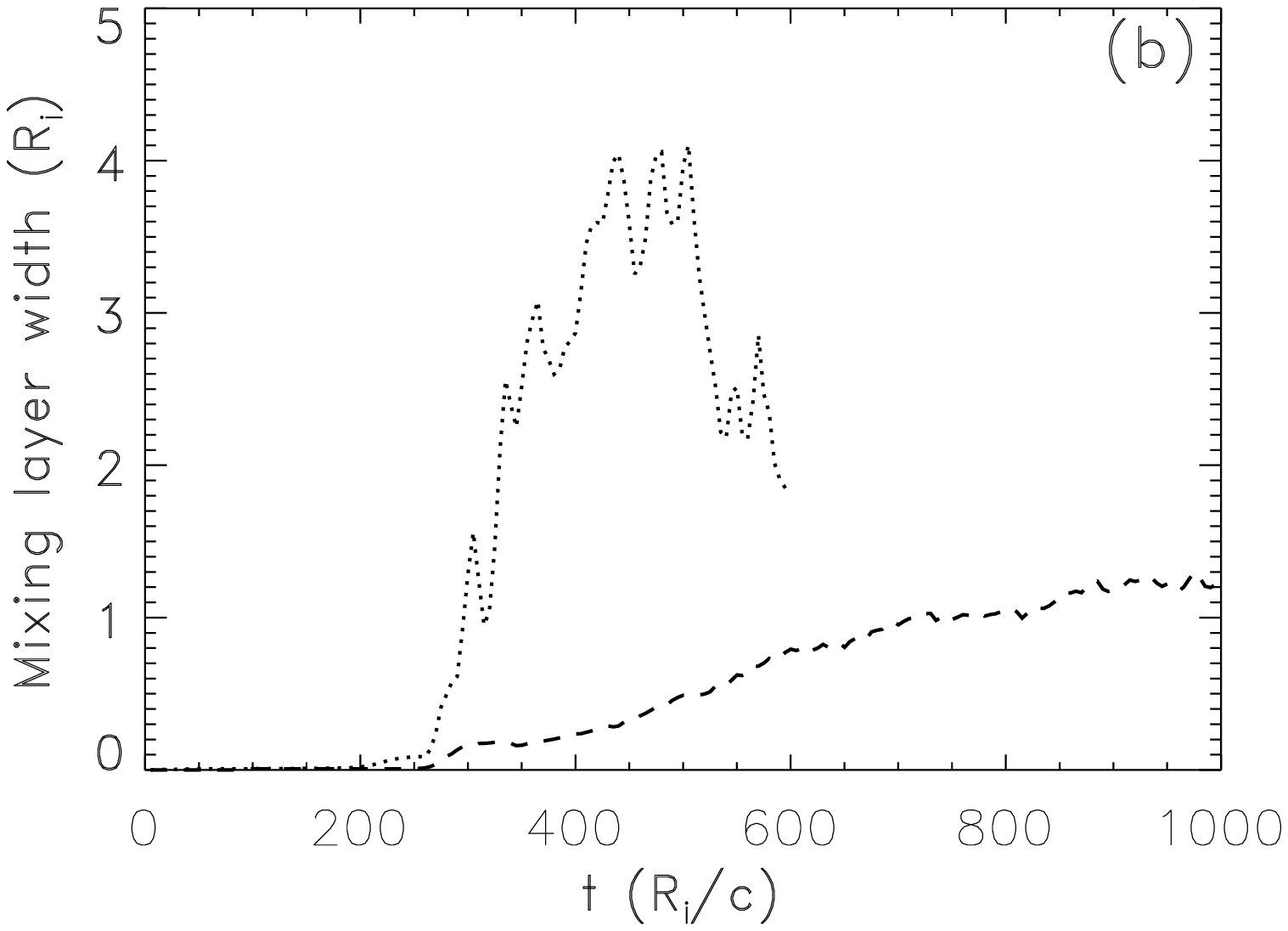}
\caption{\label{fig:f8} Left panel (a) shows the evolution of the
total longitudinal momentum, normalized to the initial value of
the simulation, as a function of time, for the vortex-sheet
analytical limit simulation (dotted line) and for the sheared jet
simulation (dashed line). The right panel (b) shows the width of
the mixing layer, measured as the radial distance between tracer
values of 0.95 and 0.05. The lines represent the same models as in
panel a.}
\end{figure*}
%
%%%%%%%%%%%%%%%%%%%%%%%%%%%%%%%%%%%%%%%%%%%%%%%%%%%%%%%%%%%%%%%%%%%%%

By analyzing the long-term simulation results we find that those
jets for which the resonant modes start to dominate early in the
simulation, do not disrupt, but instead widen and develop a thick
long-standing layer of very large specific internal energy. An
example of this behaviour is shown in Fig.~\ref{fig:f6} were we
show panels corresponding to the pressure, jet mass fraction
(tracer) specific internal energy and flow Lorentz factor for
Model D20 once an asymptotic quasi-steady state has been reached.
For comparison, Fig.~\ref{fig:f7} shows the equivalent set of
panels to those in Fig.~\ref{fig:f6} for the vortex-sheet
approximation case ($m=50$). Morphological and quantitative
differences, as entrainment and jet disruption, are clearly
observed. We thus find that these resonant modes shield jets
against disruption. The presence of the hot boundary layer as well
as the shear-layer resonant modes characterized by short radial
wavelengths modify the interaction of the long-wavelength sound
waves with the jet boundaries. Other facts pointing towards the
non-linear stabilizing role of the shear layer resonant modes are
shown in Fig~\ref{fig:f8}, where the evolution of the normalized
total longitudinal momentum and the width of the mixing layer are
shown as a function of time. At the end of the simulation (at time
$t= 1000 R_j/c$, well inside the non-linear regime) Model D20,
with $m=25$, has transferred less than 4\% of the axial momentum
to the ambient medium, while in the corresponding vortex sheet
case it has transferred as much as 40\% of the axial momentum at
time $t= 595 R_j/c$ (see Figs.~\ref{fig:f6} and \ref{fig:f7}). The
width of the mixing layer developed by this model in both vortex
sheet limit and sheared flow cases also points to the stabilizing
role of the shear-layer resonant modes. Whereas in the vortex case
the mixing layer grows radially up to $4\,R_j$ with an expansion
velocity $\sim\,0.01\,c$, in the sheared case it only develops up
to $1.2\,R_j$ and with a much smaller expansion velocity
$\sim\,1.2\,10^{-3}\,c$. The width of the mixing layer is computed
as the distance between the outermost radius where the tracer (jet
mass fraction) value is 0.95 and the innermost radius where its
value is 0.05. Thus, the fall in the width of the mixing layer for
the vortex case (dotted line in the plot) at the latest times of
the simulation ($t=500-600\,R_j/c$) is not due to a real reduction
of this width, but to the fact that there are portions of pure jet
material (where the tracer value is 1) moving at large radii and
close to regions of the grid where the external medium material
prevails (tracer $\approx 0$), as can be seen in
Fig.~\ref{fig:f7}. This is just an artifact of the way in which
the width of the mixing layer is computed.

\section{Discussion}

\subsection{Nature of resonant modes}
   We have analyzed under general conditions the effect of shear on the
stability properties of relativistic flows. The linear analysis
has allowed us to discover resonant modes specific to the
relativistic shear layer that have the largest growth rates. These
modes are found to develop in high Lorentz factor and relativistic
Mach number jets. The effects of the growth of these modes in the
non-linear stability of relativistic flows have been probed by a
series of high-resolution hydrodynamical simulations.

Fourier analysis of the results of the numerical simulation shows
that the fastest growing mode corresponds to the one expected from
the linear analysis. The growth rates found in the simulations are
of the order of those predicted by linear theory close to the jet
axis, but larger by factors ranging from 1.4 to 2.0 in the shear
layer, depending on the jet parameters, than those predicted by
the solutions to the linear stability problem, which might be due
to non-linear interactions between the perturbation waves in the
shear layer.

Urpin \cite{Ur02} has studied the growth of instabilities in
sheared jets. In \cite{Ur02} an analytical approach was done for
the case of cold fluid jets with a velocity shear. One of the most
important conclusions derived from that work is that the
shear-layer instabilities found may grow faster than the KH
instabilities in the vortex sheet approximation (this fact was
first pointed out by \cite{Bi91}). The similarities between the
instabilities reported in this paper and those studied by Urpin
\cite{Ur02} are found to be: 1) The growth rates are larger for
hotter jets, 2) the growth rates decrease for faster jets, and 3)
these instabilities are dominant for higher order modes. The
resonant modes reported in this paper represent a manifestation of
the so-called shear-driven instabilities, which were also reported
by Urpin \cite{Ur02} for a specific set of physical conditions in
the jet. However, the work reported in the present paper includes
a wider set of jet parameters and the support of the results of
numerical simulations and solutions (found via numerical methods)
of the differential equation of pressure perturbation. The latter
permits a deeper analysis of the linear phase of growth of the
instabilities. Also, the method developed in this paper is valid
for any shape of the shear layer.

\subsection{Formation of hot layers}
 The formation of a hot boundary layer surrounding the inner core
of the jet as a consequence of the growth of resonant modes has
been reported in the previous section on the non-linear regime. In
this section, the formation of such hot boundaries is explained.

%%%%%%%%%%%%%%%%%%%%%%%%%%%%%%%%%%%%%%%%%%%%%%%%%%%%%%%%%%%%%%%%%%%%%
%
%Fig.8

\begin{figure}[!h]
\includegraphics[width=0.48\columnwidth]{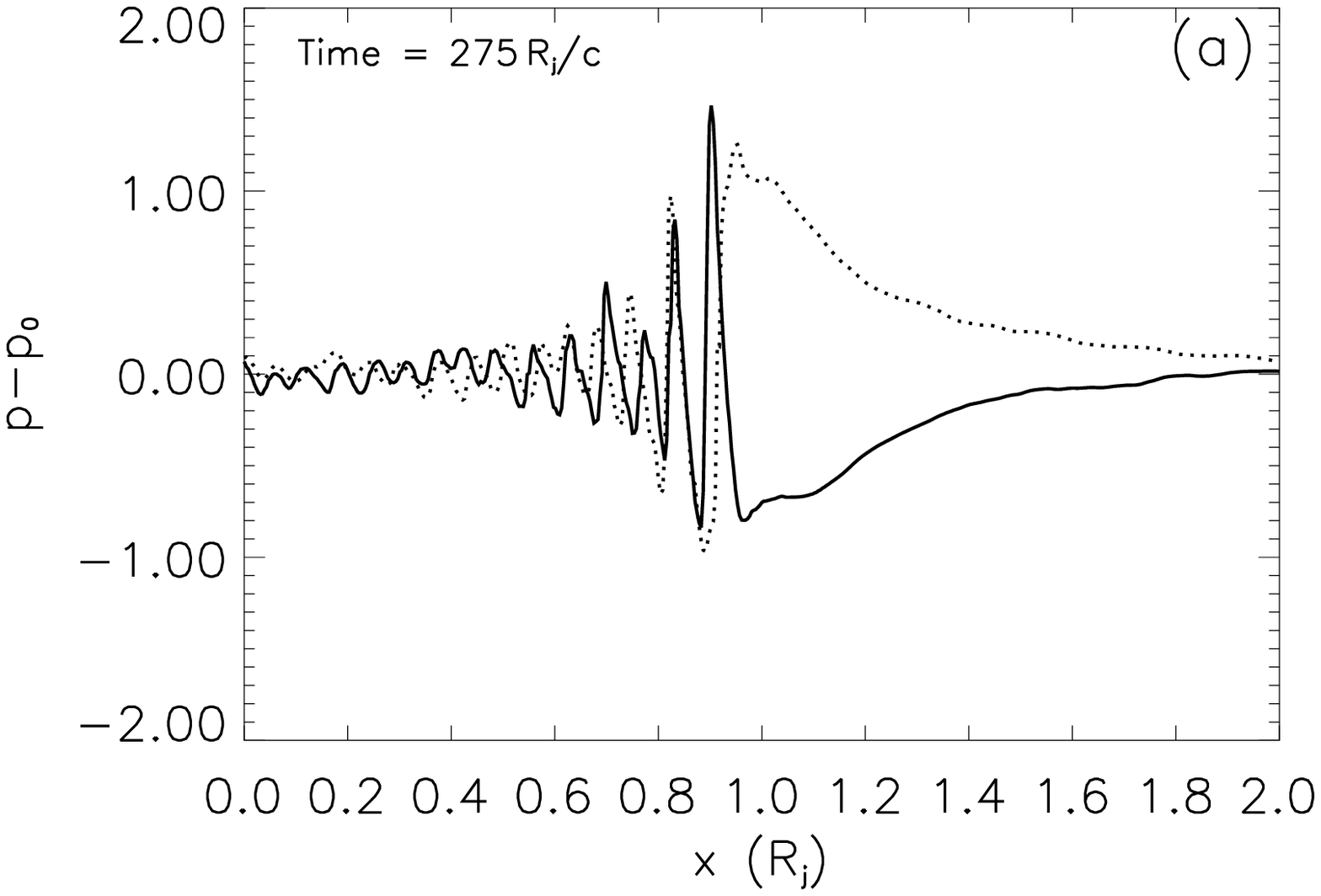}
\includegraphics[width=0.48\columnwidth]{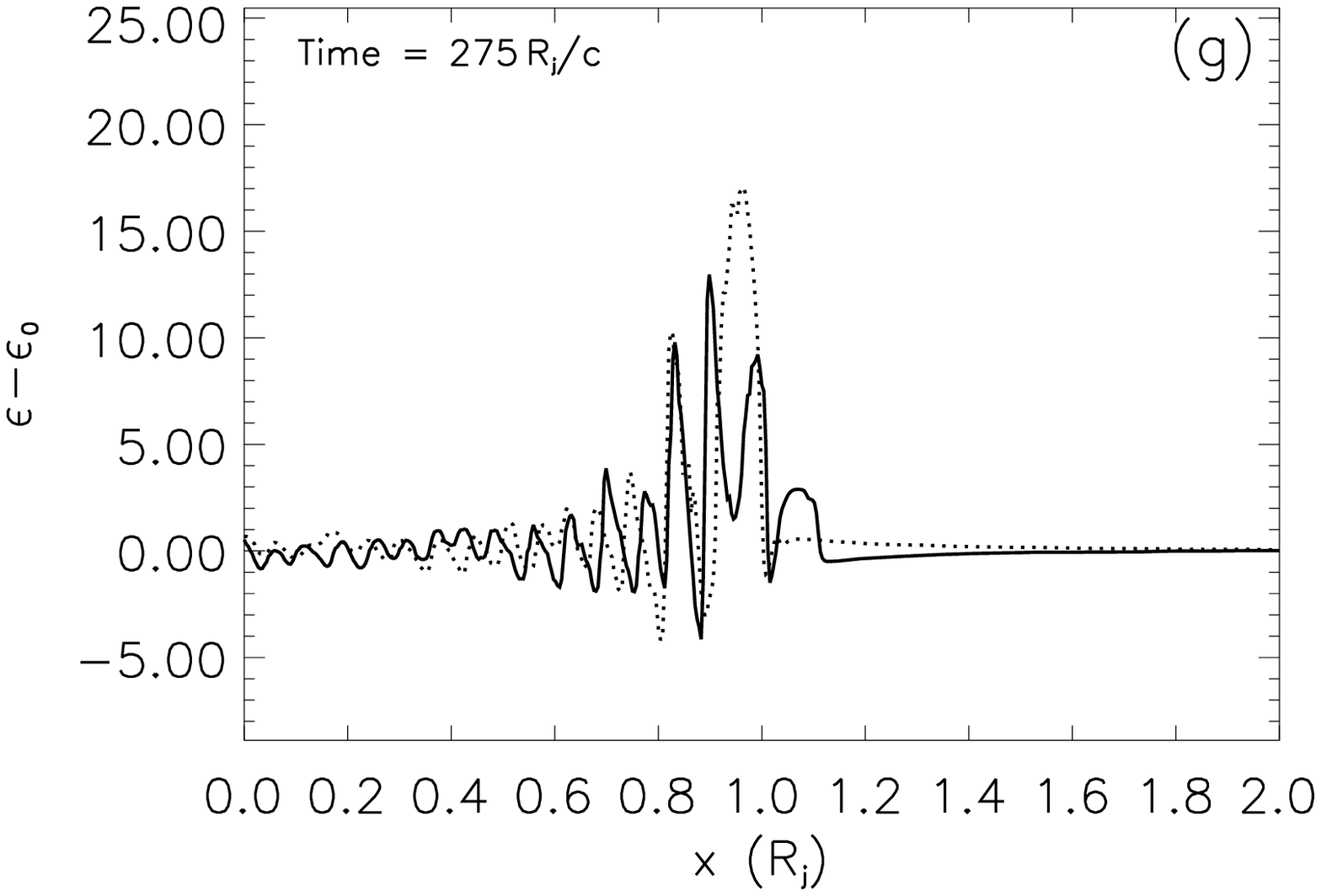}

\includegraphics[width=0.48\columnwidth]{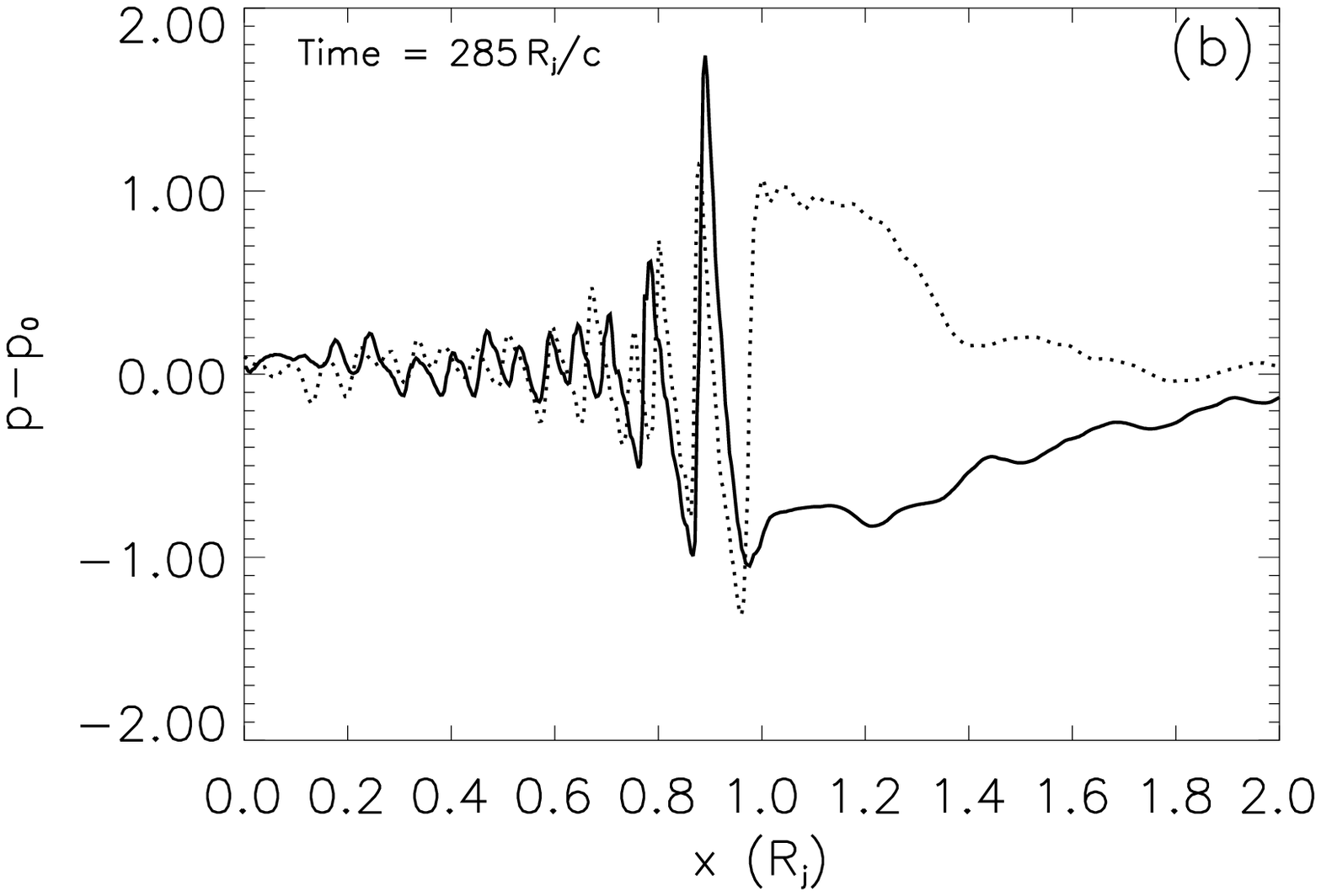}
\includegraphics[width=0.48\columnwidth]{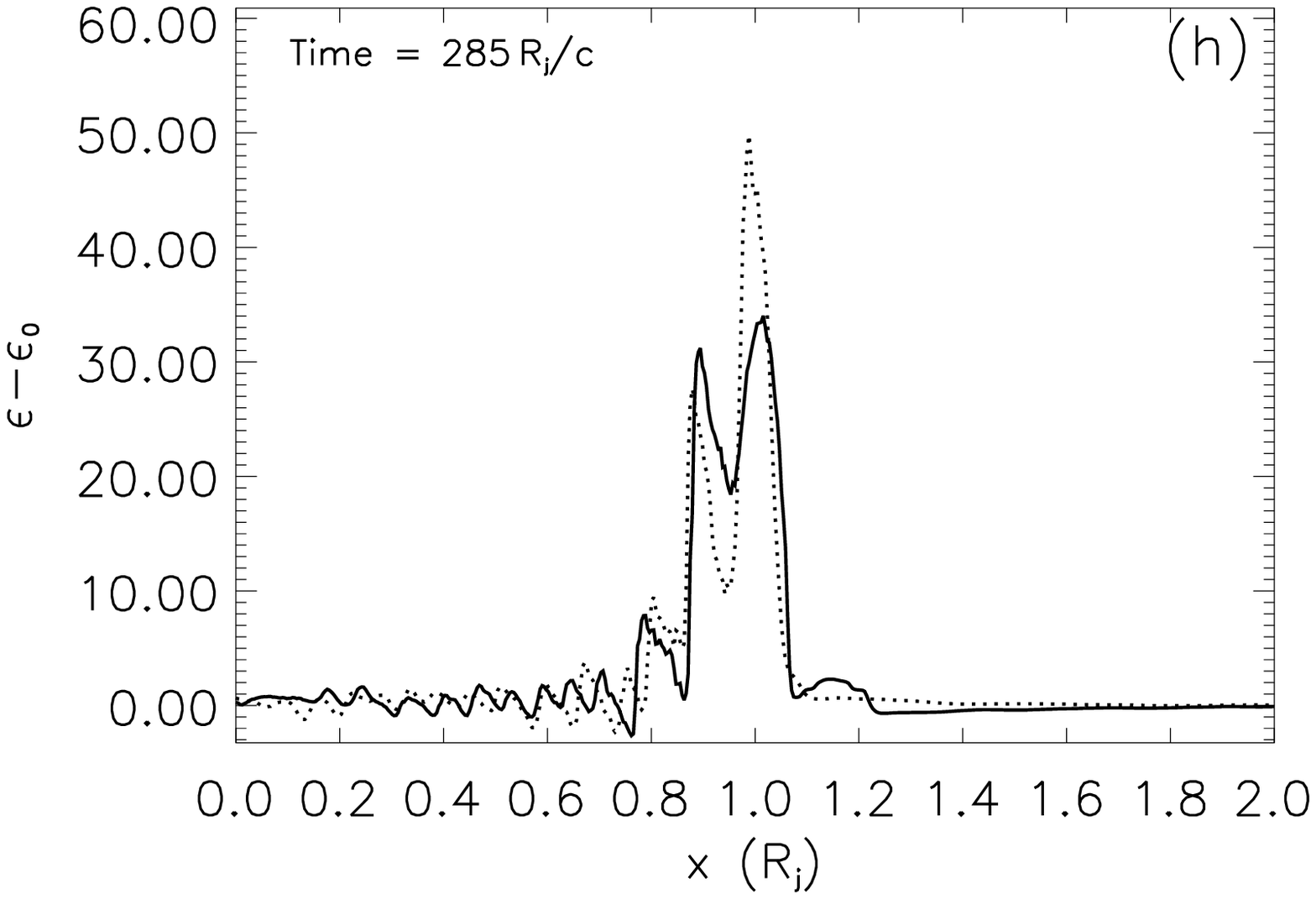}

\includegraphics[width=0.48\columnwidth]{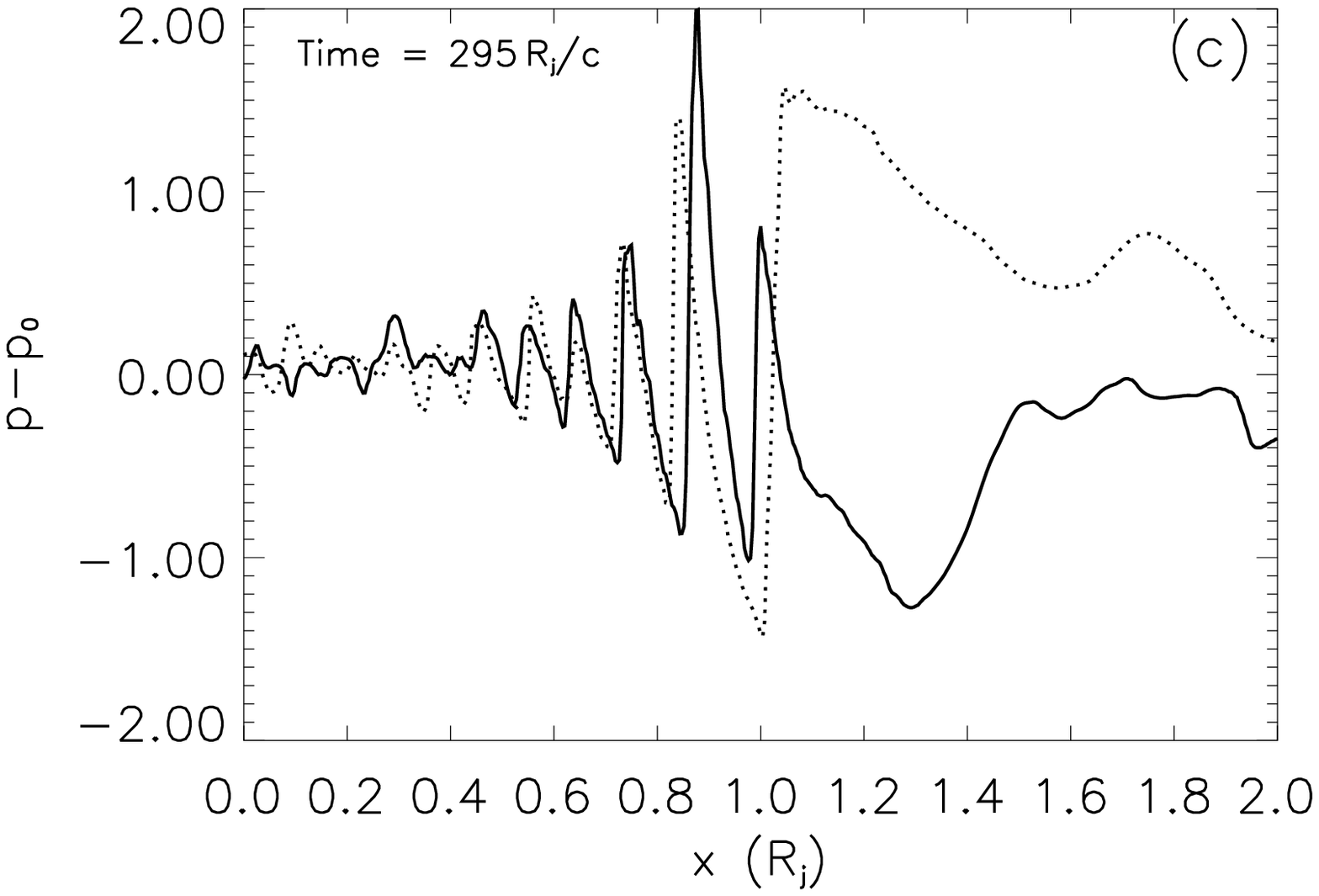}
\includegraphics[width=0.48\columnwidth]{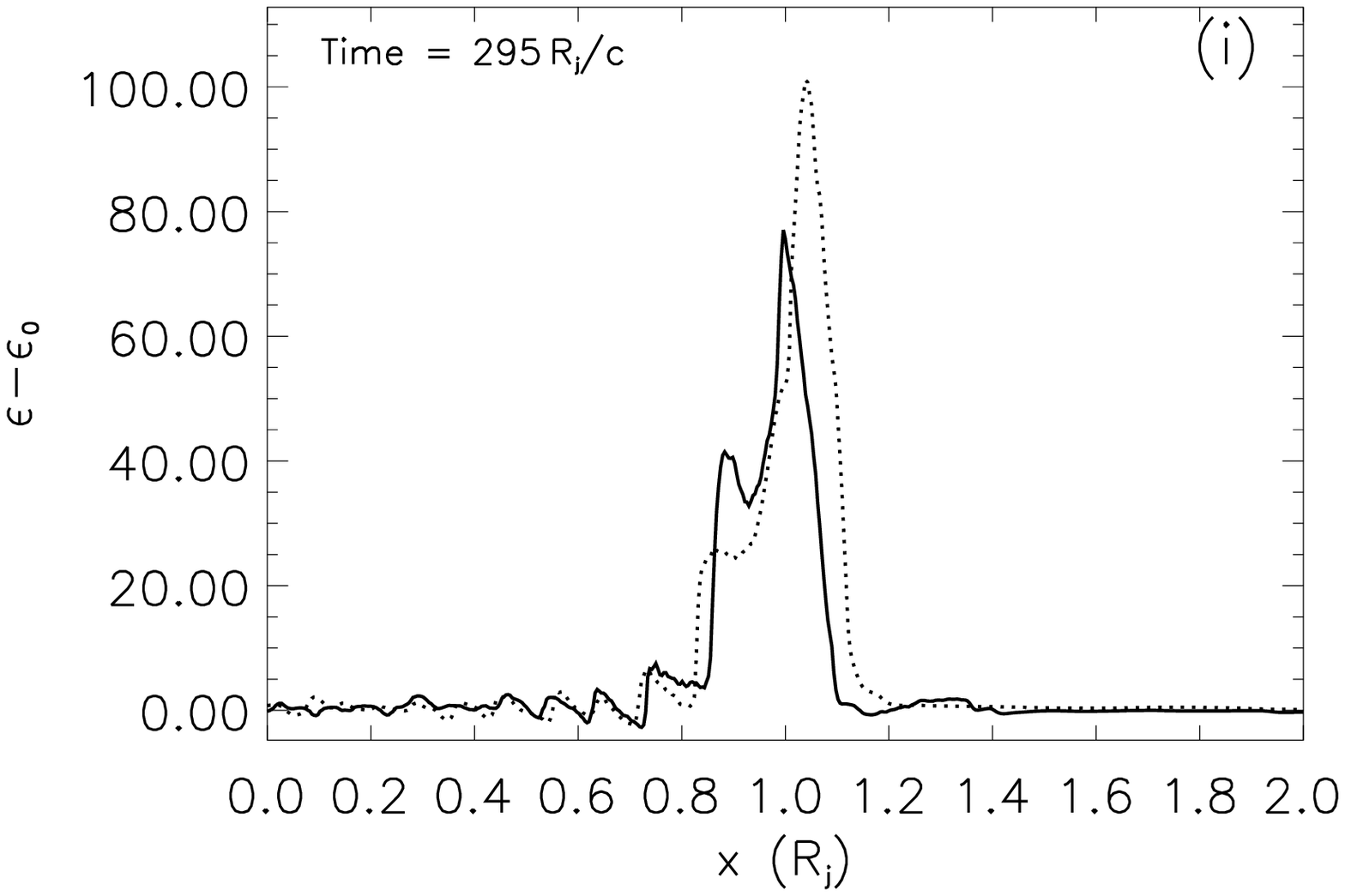}

\includegraphics[width=0.48\columnwidth]{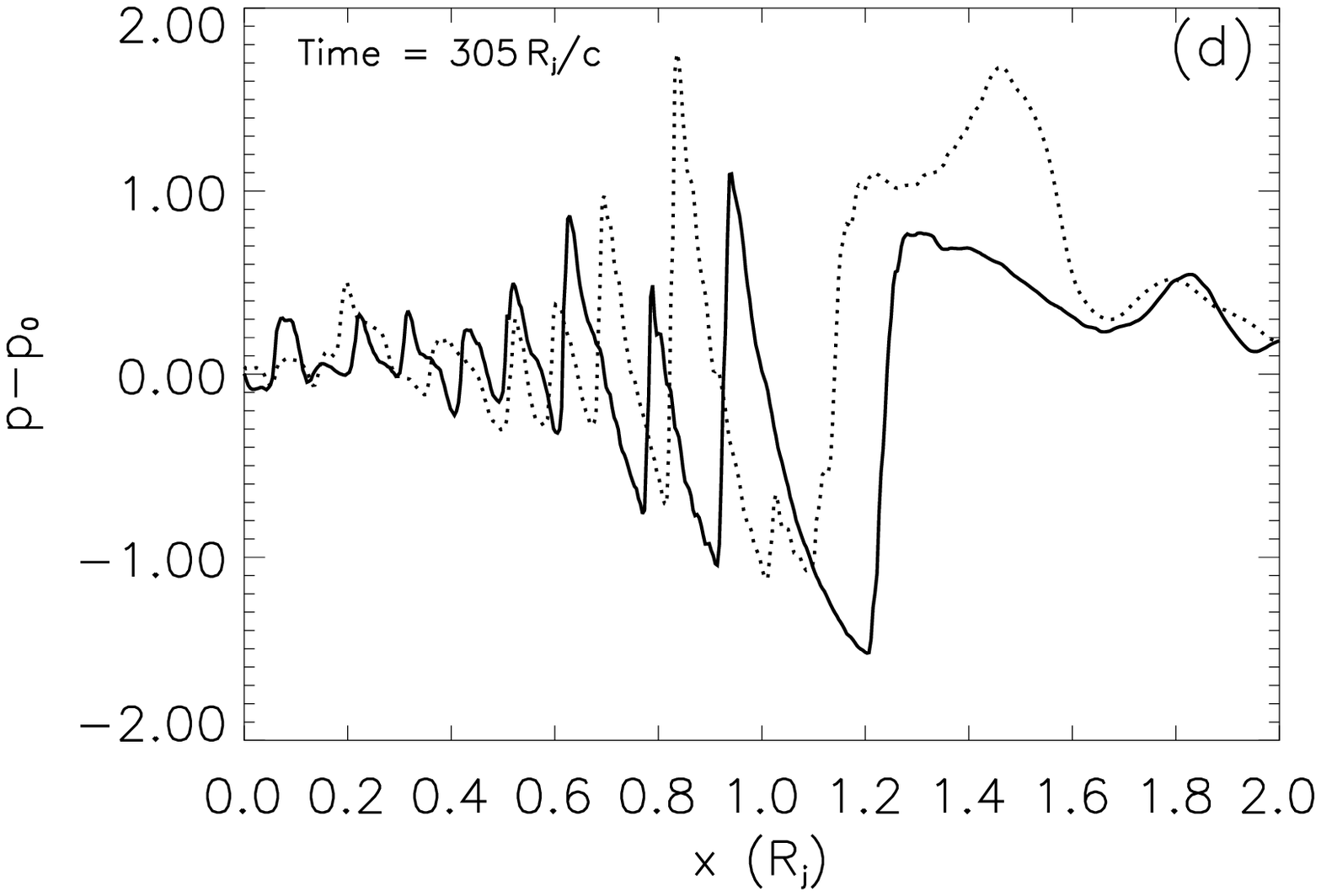}
\includegraphics[width=0.48\columnwidth]{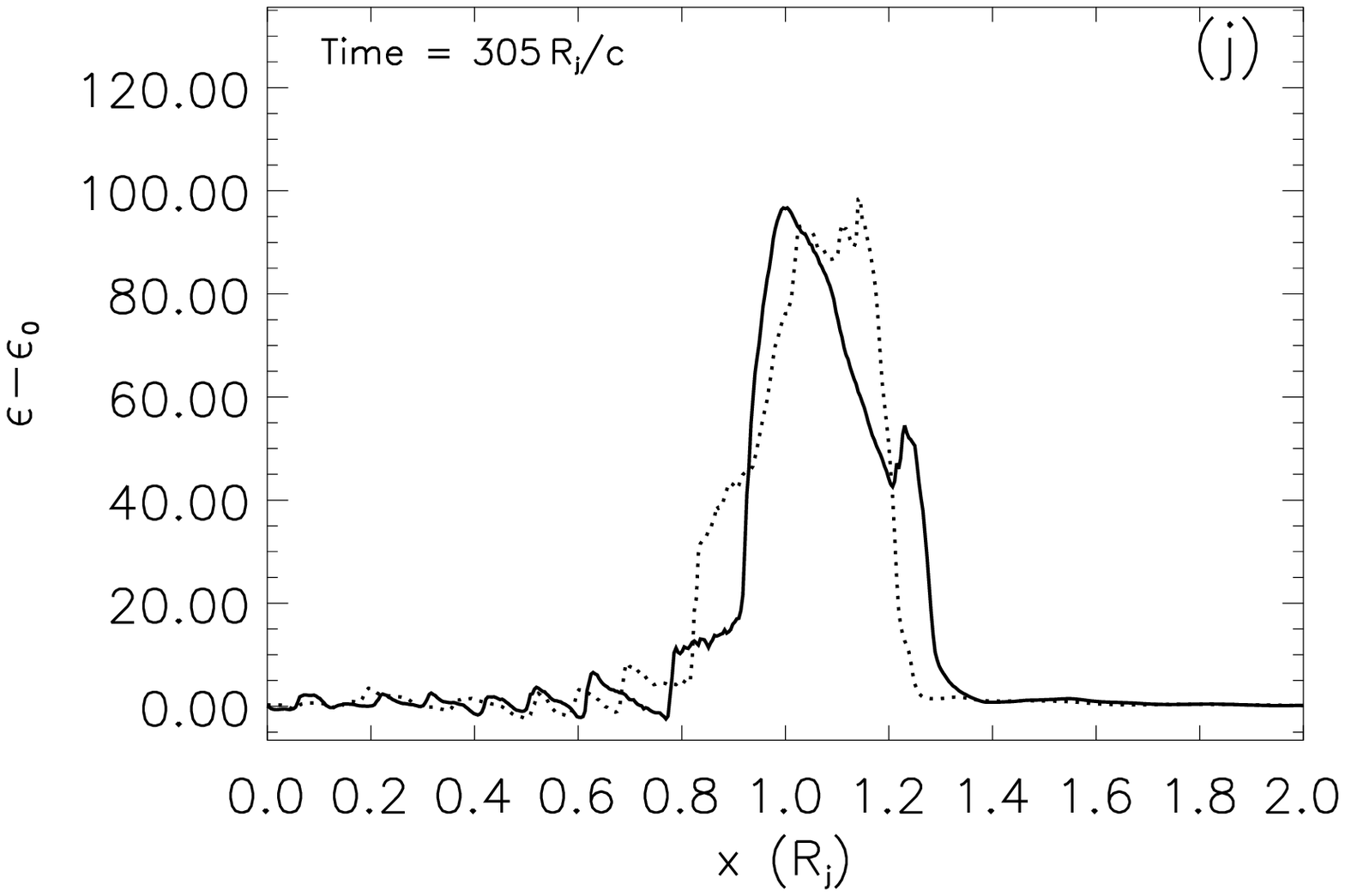}

\includegraphics[width=0.48\columnwidth]{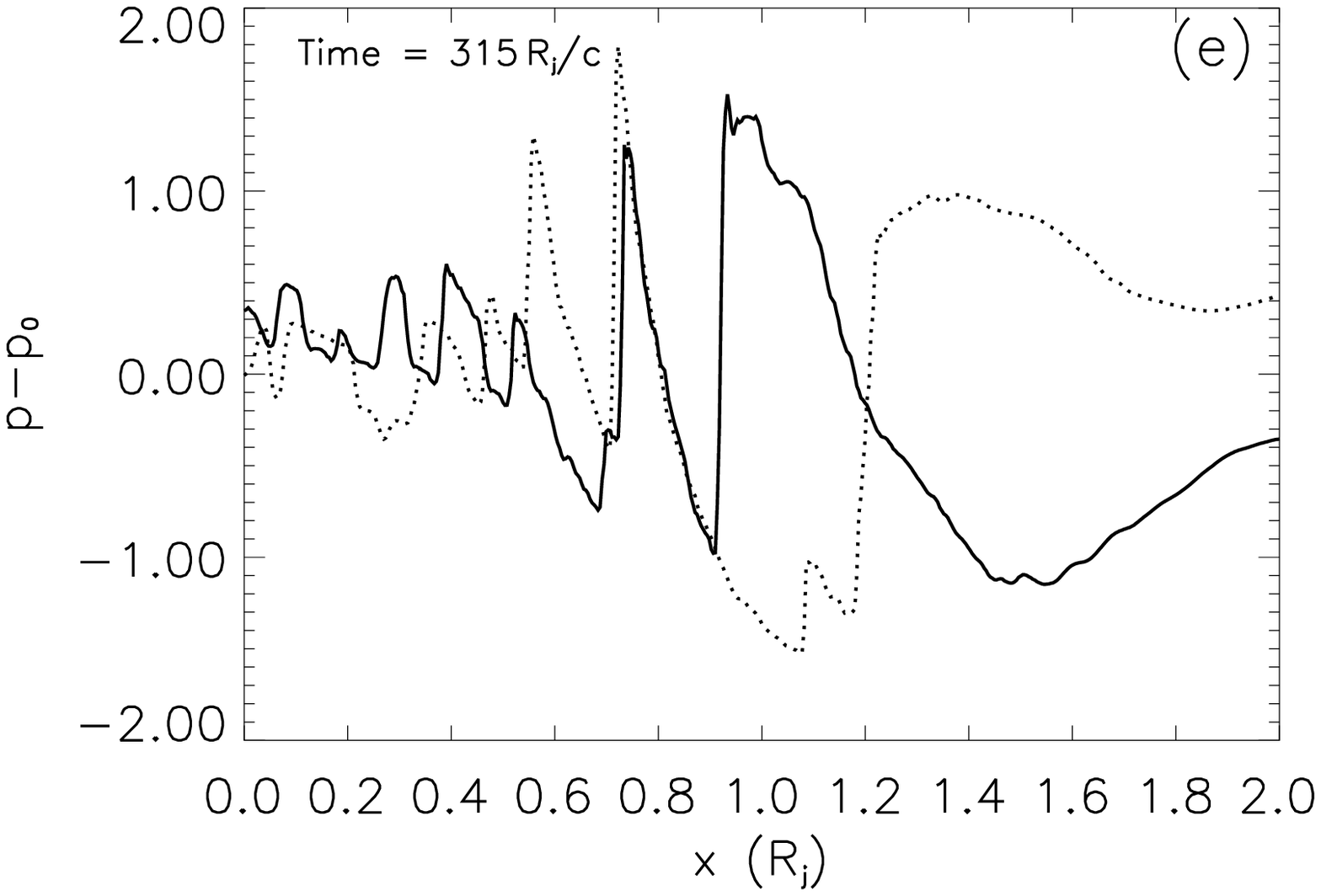}
\includegraphics[width=0.48\columnwidth]{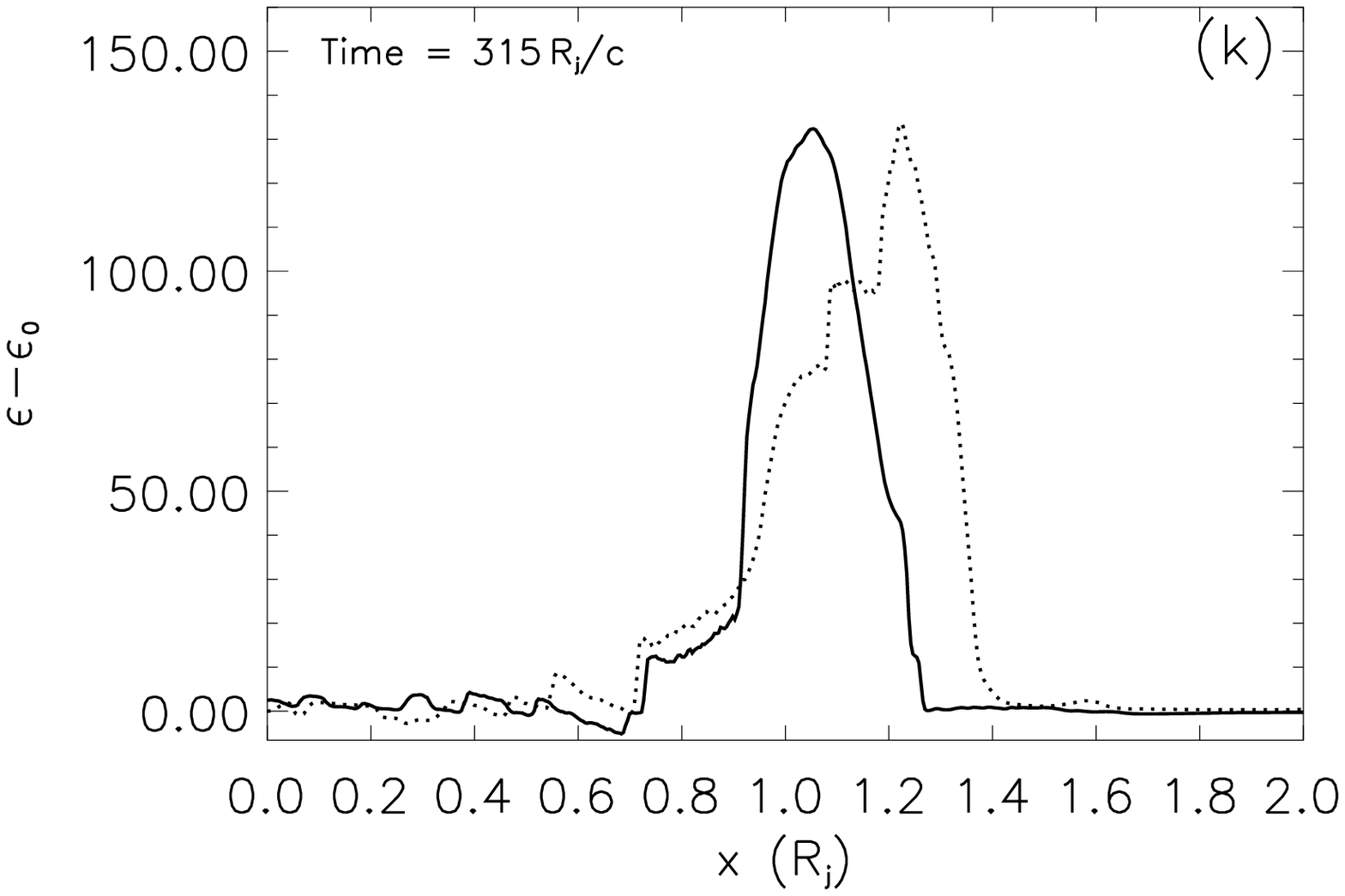}

\includegraphics[width=0.48\columnwidth]{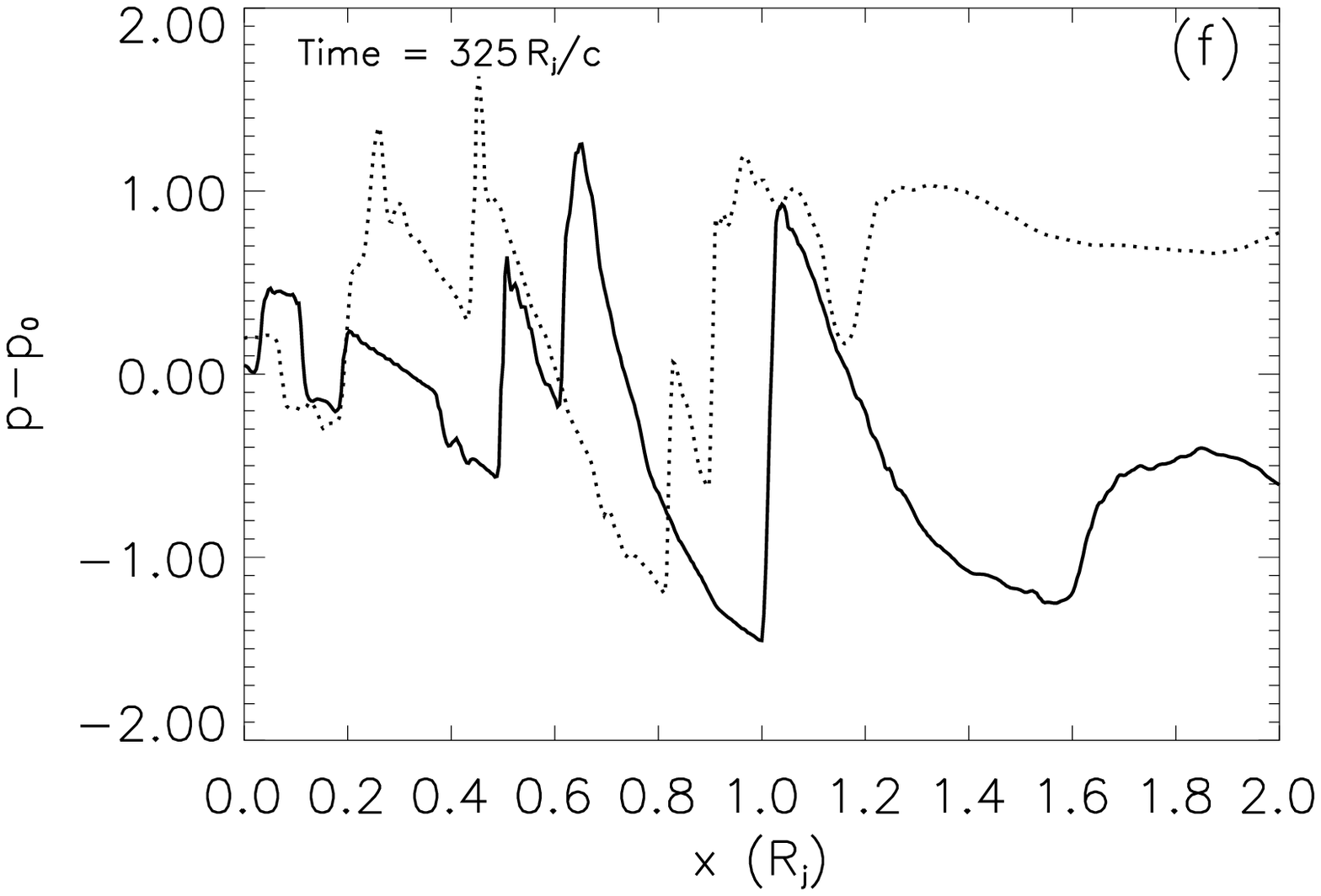}
\includegraphics[width=0.48\columnwidth]{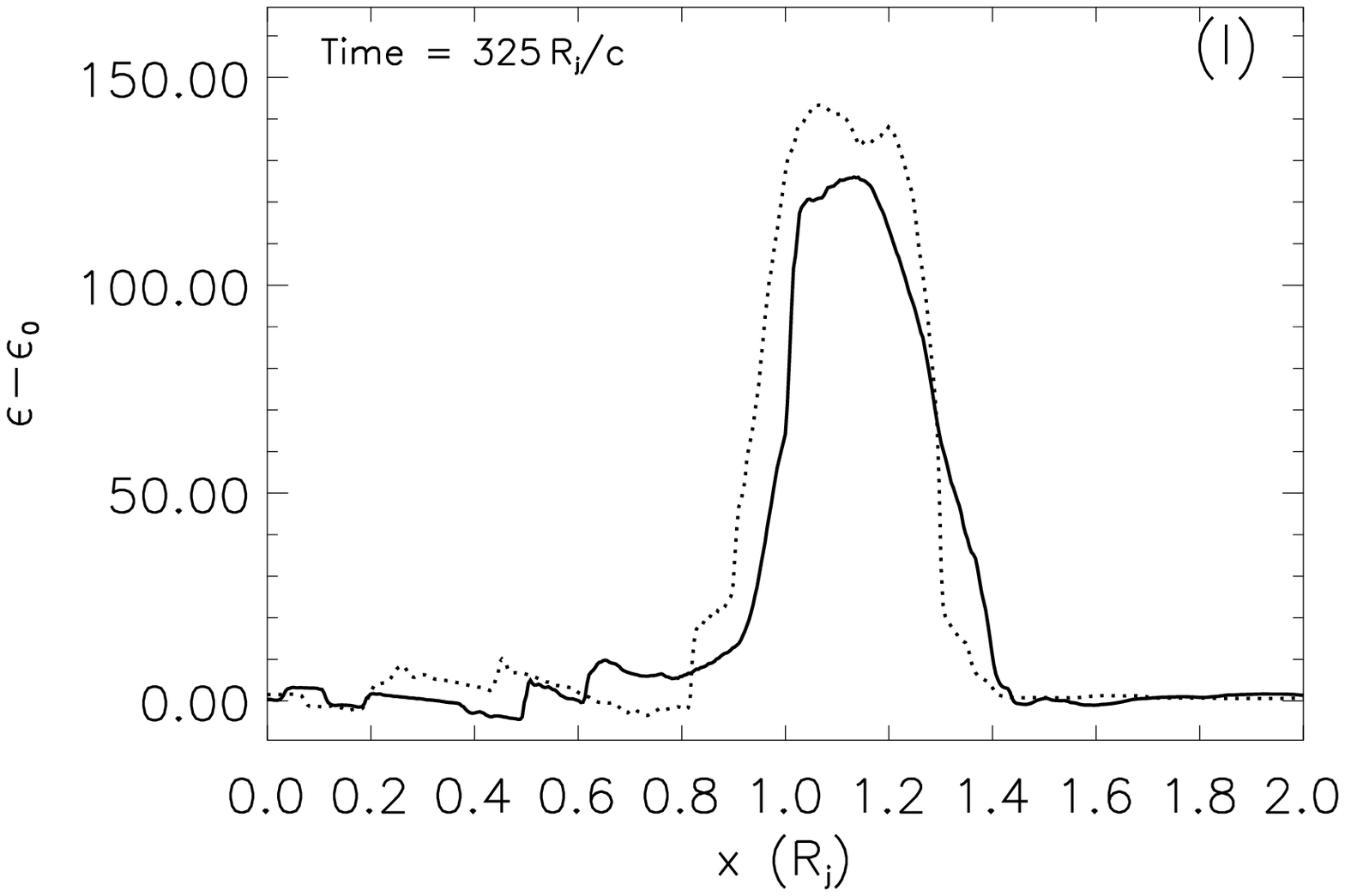}

\caption{\label{fig:f9} Radial plots of pressure perturbation
($P-P_0$, with $P_0=2.0\, \rho_{ext}\,c^2$, panels a-f) and
specific internal energy ($\varepsilon-\varepsilon_0$, with
$\varepsilon_0=60.0 c^2$, panels g-l) at different times in
simulation of Model D20. Solid line stands for pressure
perturbation at $z=0\,R_j$ and dotted line stands for the pressure
perturbation at half grid $z=4\,R_j$. The plots show how the
steepening of the pressure waves and dissipation in shocks leads
to heating of the shear layer. Note the different scales (increase
of the maxima with time) for the specific internal energy
perturbation plots.}

\end{figure}
%
%%%%%%%%%%%%%%%%%%%%%%%%%%%%%%%%%%%%%%%%%%%%%%%%%%%%%%%%%%%%%%%%%%%%%

 The parallel and perpendicular wavelengths of the shear-layer
resonant modes, $\lambda_z$ and $\lambda_x$, respectively, are
both small ($\alt R_j$) with $\lambda_x \ll \lambda_z$. Therefore
their wavevectors are almost perpendicular to the jet axis and
thus the waves propagate from the shear layer towards the jet
axis. On the other hand the resonant modes have large growth
rates, exceeding the growth rate of other modes, so they start to
dominate the evolution. In \cite{PH04} it was shown that the
growth of instabilities goes through three main stages: linear
phase, saturation phase and non-linear phase. The saturation of
the linear growth of KH instabilities in relativistic flows is
stopped when the amplitude of velocity perturbation reaches the
speed of light in the jet reference frame. As the maximum
amplitude is reached, the sound waves propagating towards the jet
axis (in the jet reference frame) steepen and form shock fronts.
The fluid particles moving outwards from the jet interior cross
the shock, decelerate and increase their internal energy. In
addition, turbulent motions of particles, as they go through
shocks and generate small scale velocity variations, also
contribute to the conversion of kinetic energy into internal
energy. Fig.~\ref{fig:f9} illustrates the process of generation of
the hot shear layer which protects the central core of the jet at
the end of the linear regime for Model D20. In the left panels
(panels a to f) we display radial plots of the pressure
perturbation at different times in the transition from the linear
to the non-linear regime. The plots show how the maxima of
pressure perturbation appear in the shear layer and how the waves
steepen. In the right panels (g to l) we display radial plots of
the perturbation in specific internal energy, and show how the
shocks produced by the steepening of the waves expand and heat the
shocked material in the shear layer.

\section{Implications for extragalactic jets}

 Our results offer an explanation to the morphological FRI/FRII dichotomy of large scale
extragalactic radio jets \cite{FR74} and its present paradigm.
This dichotomy consists on a morphological classification of
extragalactic jets, being FRII sources those showing a high
collimation and bright hot-spot in the point of collision with the
ambient, and FRI sources those showing a diffuse and decollimated
morphology in their outer regions. The latter has been interpreted
as due to jet disruption and mass loading of the original flow
\cite{LB02}. The growth of the shear layer resonances in the
highly relativistic models considered in this paper, can explain
the remarkable collimation and stability properties of powerful
radio jets. Current theoretical models \cite{LB02} interpret FRI
morphologies as the result of a smooth deceleration from
relativistic ($\gamma \leq 3$, \cite{Pe96}) to non-relativistic
transonic speeds ($\sim 0.1\,c$) on kpc scales. On the contrary,
radio-flux asymmetries between jets and counter-jets in the most
powerful radio galaxies and quasars (FRII) indicate that
relativistic motion ($\gamma\sim 2-4$, \cite{Br94}) extends up to
kpc scales in these sources. In addition, current models for high
energy emission from powerful jets at kpc scales \cite{Ce01} offer
additional support to the hypothesis of relativistic bulk speeds
on these scales. This whole picture is in agreement with the
results presented here as the development of resonant, stabilizing
modes occur in faster jets, while slower jets appear to be
disrupted by entrainment of ambient material and slowed down to
$v<0.5\,c$ during their evolution. These conclusions point to an
important contribution by intrinsic properties of the source to
the morphological dichotomy. Nevertheless, the importance of the
ambient medium cannot be ruled out on the basis of our
simulations, since we consider an infinite jet in pressure
equilibrium flowing in an already open funnel and surrounded by a
homogeneous ambient medium.

 There are plenty of arguments indicating the existence of
transversal structure in extragalactic jets at all scales
\cite{LZ01,Pe05,Sw98}. We have found the development of relatively
thin ($\approx 2 R_j$), hot shear layers in models affected by the
growth of resonant modes to nonlinear amplitudes, as discussed in
this paper. These hot shear layers could explain several
observational trends in the transversal structure of powerful jets
at both parsec and kiloparsec scales \cite{Sw98}. Conversely and
according to our simulations, these transition layers could be
responsible for the stability of fast, highly supersonic jets,
preventing the mass-loading and subsequent disruption. Thicker,
mixing layers formed in slower jets could mimic the transition
layers invoked in models of FRIs \cite{LB02}.

 Direct comparison of our results with real jets is however still
difficult due to the slab geometry of the problem studied here and
to the fact that magnetic fields are not considered in our work.
The latter are known to be present in extragalactic jets and even
to be dynamically important for the evolution of compact jets.
Several authors have studied their influence on the stability
these objects (\cite{ha06}). The inclusion of magnetic fields and
three dimensional cylindrical geometries in linear calculations
and numerical simulations is a natural further step in our work.

\begin{acknowledgments}
Calculations were performed in SGI Altix 3000 computer {\it CERCA}
at the Servei d'Inform\`atica de la Universitat de Val\`encia.
This work was supported by the Spanish DGES under grant
AYA-2001-3490-C02 and Conselleria d'Empresa, Universitat i Ciencia
de la Generalitat Valenciana under project GV2005/244. M.P.
benefited from a predoctoral fellowship of the Universitat de
Val\`encia ({\it V Segles} program) and a postdoctoral fellowship
in the Max-Planck-Institut f\"ur Radioastronomie in Bonn.
\end{acknowledgments}

%\newpage %Just because of unusual number of tables stacked at end
\bibliography{apssamp.bib}% Produces the bibliography via BibTeX.

\begin{thebibliography}{}

\bibitem[1]{Ch61} S. Chandrasekhar, {\it Hydrodynamic and
hydromagnetic stability}, Clarendon Press 1961; A.E. Gill,
Phys. Fluids {\bf 8}, 1428 (1965); R.A. Gerwin, Rev. Mod. Phys. {\bf
40}, 652 (1968)

\bibitem[2]{TS76} B.D. Turland and P.A.G. Scheuer, MNRAS {\bf 176},
421 (1976); R.D. Blandford and J.E. Pringle, MNRAS {\bf 176}, 443
(1976)

\bibitem[3]{FT78} A. Ferrari et al., Astron. Astrophys. {\bf 64}, 43
(1978)

\bibitem[4]{Ha79} P. Hardee, Astrophys. J. {\bf 234}, 47 (1979)

\bibitem[5]{Ha87a} P. Hardee, Astrophys. J. {\bf 313}, 607 (1987)

\bibitem[6]{Ha87b} P. Hardee, Astrophys. J. {\bf 318}, 78 (1987)

\bibitem[7]{comment2} e.g., P.E. Hardee, Astrophys. J. {\bf 250}, 9
(1981); P.E. Hardee and M.L. Norman, Astrophys. J. {\bf 342}, 680
(1989); J.-H.Zhao et al., Astrophys. J. {\bf 387}, 69 (1992)

\bibitem[8]{comment3} Agudo et al., Astrophys. J. Lett. {\bf 549}, 183
  (2001): pinching KH modes generate radio knots with distinct
  kinematical properties. Hardee et al., Astrophys. J. {\bf 555}, 744
  (2001): instabilities forced by precession and wave-wave
  interactions can explain differentially moving features in the
  jets. Walker et al. 2001, Astrophys. J. {\bf 556}, 756 (2001):
  interpretation of the structure and motions of the 3C120 radio jet
  (0.6-300 pc). Lobanov et al., New Astron. Rev. {\bf 47}, 629 (2003):
  internal structure and dynamics of the M87 jet.

\bibitem[9]{LZ01}A.P. Lobanov and J.A. Zensus, Science {\bf 294}, 128
  (2001)

\bibitem[10]{BD75}W. Blumen et al., J. Fluid Mech. {\bf
    71}, 305 (1975); P.G. Drazin and A. Davey, A., J. Fluid Mech. {\bf
    82}, 255 (1977)

\bibitem[11]{FM82}A. Ferrari et al., MNRAS {\bf 198}, 1065 (1982)

\bibitem[12]{Bi91}M. Birkinshaw, MNRAS {\bf 252}, 505 (1991)

\bibitem[13]{HS96}M. Hanasz and H. Sol, Astron. Astrophys. {\bf 315},
  355 (1996)

\bibitem[14]{Ur02}V. Urpin, Astron. Astrophys. {\bf 385}, 14 (2002)

\bibitem[15]{comment4a} A. Rosen et al., Astrophys. J. {\bf 516}, 729
  (1999); P.E. Hardee, Astrophys. J. {\bf 533}, 176 (2000);
  P.E. Hardee et al., Astrophys. J. {\bf 555}, 744 (2001)

\bibitem[16]{comment4b} M.A. Aloy et al., Astron. Astrophys. {\bf
396}, 693 (2002)

\bibitem[17]{Pe05} M. Perucho, J.~M. Mart\'{\i}, M. Hanasz, A\&A, {\bf 443},
863 (2005)

\bibitem[18]{Bi84}The equation was first derived by M. Birkinshaw,
MNRAS {\bf 208}, 887 (1984)

\bibitem[19]{RL84}S. Roy Choudhury and R.V.E. Lovelace,
Astrophys. J. {\bf 283}, 331 (1984)

\bibitem[20]{Press97}
W.H. Press et al., {\it Numerical recipes}, Cambridge Univ. Press
1997

\bibitem[21]{PH04} The notation for the models follows that of
M. Perucho et al., Astron. Astrophys. {\bf 427}, 415 (2004a) and
M. Perucho et al., Astron. Astrophys. {\bf 427}, 431 (2004b) were the
linear and non-linear stability of relativistic planar jets in the
vortex-sheet case was investigated.

\bibitem[22]{comment5}Some resonances due to the presence of a shear
layer \cite{Bi91} or a sheath surrounding relativistic jet
component \cite{HS96} could appear as oscillations ({\it ripples})
on the growth rate curve in the large wave number limit, however
we do not observe oscillations of this type in our case. These
ripples could be related to the presence of discontinuities in the
jet/ambient transition, to the shape of the shear layer, or
associated to a different region in the parameter space.

\bibitem[23]{MM97} Numerical simulations were performed using a
finite-difference code based on a high-resolution shock-capturing
scheme which solves the equations of relativistic hydrodynamics
written in conservation form (J.M. Mart\'{\i} et al.,
Astrophys. J. {\bf 479}, 151 (1997)). The code was recently
parallelized using OMP directives.


\bibitem[24]{FR74} B.L. Fanaroff and J.M. Riley, MNRAS, {\bf 167}, 31 (1974)

\bibitem[25]{LB02} R.A. Laing and A.H. Bridle, MNRAS, \textbf{336},
328 (2002); R.A. Laing and A.H. Bridle, MNRAS, \textbf{336}, 1161
(2002)

\bibitem[26]{Pe96} T.J. Pearson, in {\em Energy Transport in Radio
Galaxies and Quasars}, Hardee, P.E., Bridle, A.H., and Zensus,
J.A., eds., 97 (1996)

\bibitem[27]{Br94} A.H. Bridle et al., AJ, {\bf 108}, 766 (1994)

\bibitem[28]{Ce01} A. Celotti and R.D. Blandford, in Proceedings of
\emph{Black Holes in Binaries and Galactic Nuclei}.  Kaper, L.,
van den Heuvel, E.P.J., Woudt, A.P., eds., 206 (2001)

\bibitem[29]{Sw98}M.R. Swain et al., Astrophys. J., {\bf 507}, L29
(1998); J.M. Attridge et al., Astrophys. J. Lett. , {\bf 518}, 87
(1999)

\bibitem[30]{ha06} See Hardee, P.E., in \emph{Relativistic
Jets}. Hughes, P.A., Bregman, J.N., eds., 57 (2006), and
references therein.

\end{thebibliography}

\end{document}